\documentclass[twocolumn,english,amsmath,amssymb,superscriptaddress,nofootinbib]{revtex4-1}
\usepackage[T1]{fontenc}
\usepackage[latin9]{inputenc}
\usepackage{float}
\usepackage{amsmath}
\usepackage{amssymb}
\usepackage{graphicx}
\usepackage{babel}

\usepackage{color}

\usepackage{ulem}

\newcommand{\stkout}[1]{\ifmmode\text{\sout{\ensuremath{#1}}}\else\sout{#1}\fi}

\begin{document}
\title{Ultrafast Plasmon-mediated Superradiance from Vertically Standing Molecules in Metallic Nanocavities}

\author{Yuan Zhang }
\email{yzhuaudipc@zzu.edu.cn}
\affiliation{Henan Key Laboratory of Diamond Optoelectronic Materials and Devices, Key Laboratory of Material Physics, Ministry of Education, School of Physics and Microelectronics, Zhengzhou University, Daxue Road 75, Zhengzhou 450052}
\affiliation{Institute of Quantum Materials and Physics, Henan Academy of Sciences, Mingli Road 266-38, Zhengzhou 450046}

\author{Yuxin Niu}
\affiliation{Department of Physics, University of Science and Technology Beijing, 100083 Beijing, People's Republic of China}

\author{Shunping Zhang}
\affiliation{School of Physics and Technology, Center for Nanoscience and Nanotechnology, and Key Laboratory of Artificial Micro- and Nano-structures of Ministry of Education, Wuhan University, Wuhan 430072, China}
\affiliation{Institute of Quantum Materials and Physics, Henan Academy of Sciences, Mingli Road 266-38, Zhengzhou 450046}

\author{Yao Zhang}
\affiliation{Hefei National Research Center for Physical Sciences at the Microscale and Synergetic Innovation Centre of Quantum Information and Quantum Physics, University of Science and Technology of China, Hefei, Anhui 230026, China}

\author{Shi-Lei Su}
\affiliation{Henan Key Laboratory of Diamond Optoelectronic Materials and Devices, Key Laboratory of Material Physics, Ministry of Education, School of Physics and Microelectronics, Zhengzhou University, Daxue Road 75, Zhengzhou 450052}
\affiliation{Institute of Quantum Materials and Physics, Henan Academy of Sciences, Mingli Road 266-38, Zhengzhou 450046}

\author{Guangchao Zheng}
\affiliation{Henan Key Laboratory of Diamond Optoelectronic Materials and Devices, Key Laboratory of Material Physics, Ministry of Education, School of Physics and Microelectronics, Zhengzhou University, Daxue Road 75, Zhengzhou 450052}
\affiliation{Institute of Quantum Materials and Physics, Henan Academy of Sciences, Mingli Road 266-38, Zhengzhou 450046}

\author{Luxia Wang}
\email{luxiawang@sas.ustb.edgu.cn}
\affiliation{Department of Physics, University of Science and Technology Beijing, 100083 Beijing, People's Republic of China}

\author{Gang Chen}
\affiliation{Henan Key Laboratory of Diamond Optoelectronic Materials and Devices, Key Laboratory of Material Physics, Ministry of Education, School of Physics and Microelectronics, Zhengzhou University, Daxue Road 75, Zhengzhou 450052}
\affiliation{Institute of Quantum Materials and Physics, Henan Academy of Sciences, Mingli Road 266-38, Zhengzhou 450046}

\author{Hongxing Xu}
\affiliation{School of Physics and Technology, Center for Nanoscience and Nanotechnology, and Key Laboratory of Artificial Micro- and Nano-structures of Ministry of Education, Wuhan University, Wuhan 430072, China}
\affiliation{Institute of Quantum Materials and Physics, Henan Academy of Sciences, Mingli Road 266-38, Zhengzhou 450046}

\author{Chongxin Shan}
\email{cxshan@zzu.edu.cn}
\affiliation{Henan Key Laboratory of Diamond Optoelectronic Materials and Devices, Key Laboratory of Material Physics, Ministry of Education, School of Physics and Microelectronics, Zhengzhou University, Daxue Road 75, Zhengzhou 450052}
\affiliation{Institute of Quantum Materials and Physics, Henan Academy of Sciences, Mingli Road 266-38, Zhengzhou 450046}




\begin{abstract}
Plasmon-mediated superradiance for molecules around metallic nanospheres was proposed ten years ago. However, its demonstration has not been achieved yet due to the experimental difficulty of positioning molecules, and the theoretical limitation to the enhanced collective rate of low excited molecules. In this Letter, we propose that the ultrafast plasmon-mediated superradiant pulses can be observed with strongly excited methylene blue molecules standing vertically inside gold nanoparticle-on-mirror nanocavities. Our simulations indicate that in this system the molecules could interact with each other via plasmon- and free-space mediated coherent and dissipative coupling. More importantly, the coherent coupling mediated by short-ranged propagating surface plasmons cancel largely the direct dipole-dipole coupling mediated by the free-space field, and the dominated dissipative coupling mediated by relatively long-ranged gap plasmons enables the ultra-fast superradiant pulses within picosecond scale. Our study opens up the possibility of studying the rich superradiant effects from the quantum emitters in a sub-wavelength volumn by engineering the plasmonic environments. 
\end{abstract}
\maketitle

\paragraph{Introduction.---}

Superradiance, i.e., collective spontaneous emission of the emitters, was first proposed by R. H. Dicke in 1954~\citep{RHDicke1954}, and was intensively investigated in 1980's and thereafter in both theories and experiments~\citep{AVAndreev1980}. This phenomenon has been observed in various systems, such as quantum dots~\citep{MScheibner} and nitrogen-vacancy centers~\citep{CBradac} in free space, atomic ensembles in macroscopic cavities~\citep{MANorcia2016,TLaske}, semiconductor quantum dots in microscopic cavities~\citep{FJahnke},  solid-state spins in microwave resonators~\citep{AAngerer,JDBreeze}, etc. Besides, the superradiance has also been explored to generate the atoms-photons entanglement~\citep{NLambert},  to realize ultra-narrow coherent radiation~\citep{DMeiser,JGBohnet}, as well as to demonstrate the high-precision frequency measurements~\citep{MANorcia2018}.

To realize the superradiance in free space, the emitters are often far apart as compared to the light wavelength so that their long-ranged collective but dissipative interaction dominates. If this is not satisfied, the excitons can be formed among the emitters due to the coherent dipole-dipole interaction, and the frequency detuning of  the laser excitation prohibits the strong excitation of emitters and the subsequent superradiance. Fortunately, this condition can be strongly relaxed by using dielectric/metallic micro- or nano-structures  to engineer the electromagnetic environment. For example, V. Pustovit et al., proposed theoretically  that the superradiance can occur for the emitters in a sub-wavelength volume by placing them near a metallic nanosphere~\citep{VNPustovit2009,VNPustovit2010}, where the coupling is mainly mediated by the localized surface plasmon instead of free-space field.

\begin{figure}[htp]
\begin{centering}
\includegraphics[scale=0.3]{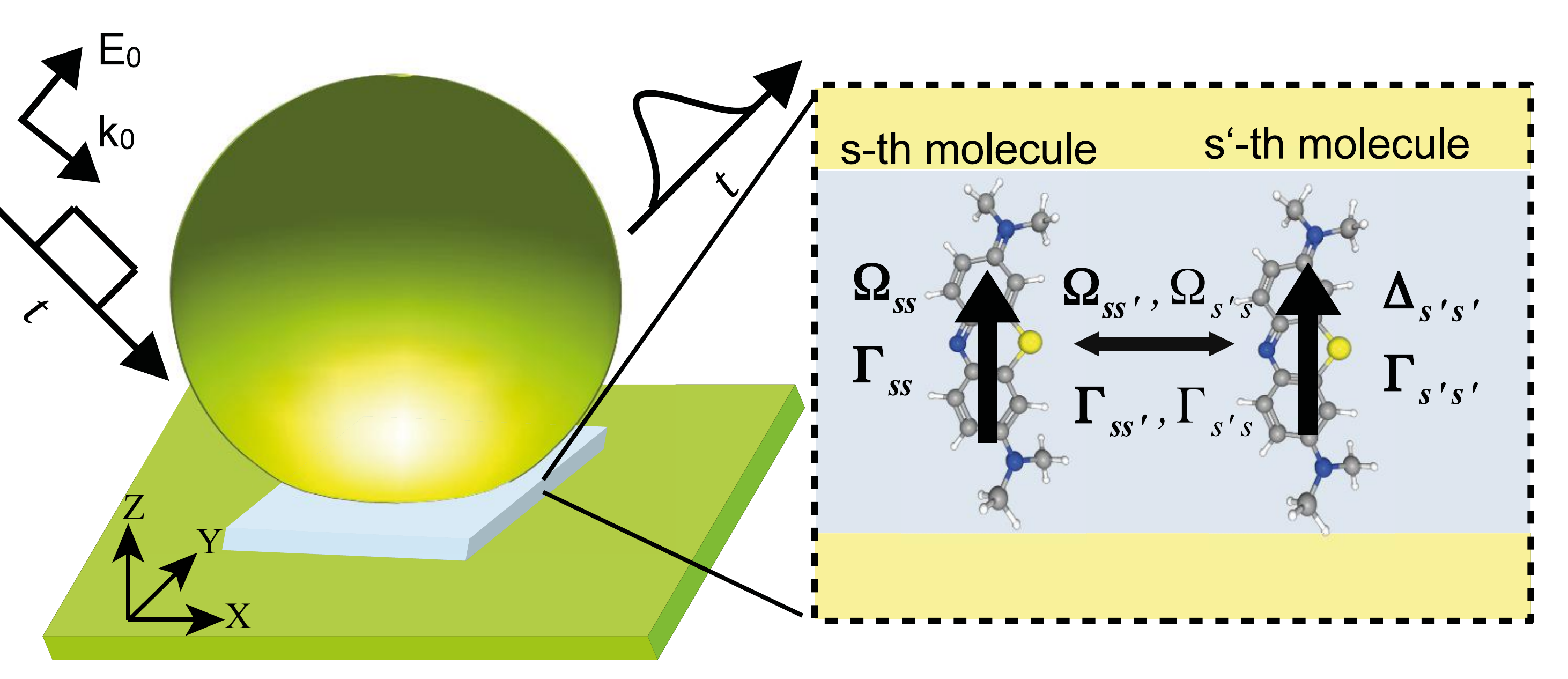}
\par\end{centering}
\caption{\label{fig:system} System schematic. Left part shows many vertically standing methylene blue molecules inside the $1.3$ nm gap of a metallic nanocavity, formed by a gold nanosphere of $90$ nm diameter truncated at bottom with a radius of $16$ nm, on a gold substrate. A laser field (with a polarization $\mathbf{E}_{0}$ and a propagation vector $\mathbf{k}_{0}$) excites the gap plasmons of the nanocavity, and  the enhanced local field excites the molecules, and finally the continuous or pulsed superradiance from the strongly excited molecules are amplified by the nanocavity as an antenna to far-field.  Right part shows the zoom-in of the gap, where the molecules (labeled by $s,s'$) are excited by the (quite) homogeneous local field (light blue), and they are shifted by $\Omega_{ss},\Omega_{s's'}$ in frequency, and dissipate with the rates $\Gamma_{ss},\Gamma_{s's'}$, as well as interact with each other through the coherent $\Omega_{ss'},\Omega_{s's}$ and dissipative couplings 
 $\Gamma_{ss'},\Gamma_{s's}$, which are mediated either by the surface plasmons or the free-space field. For more details, see text.}
\end{figure}

Although the above proposal was suggested ten years ago, plasmon-mediated superradiance has not been demonstrated yet in the experiments. Lou et al. reported the electrically driven single photon superradiance from horizontally laying molecules in a STM-based plasmonic nanocavity~\citep{YLuo}. However, this phenomenon can be largely interpreted by the formation of Franck-excitons because the typical dipole-dipole interaction dominates. The lacked verification of the proposal might be attributed to the difficulty of positioning the emitters around the metallic nanospheres, and the limitation of the theoretical model to the enhanced collective rate of the weakly excited emitters, which is also captured by a classical theory. To overcome these problems, in this Letter, we propose to investigate the superradiant pulses, as predicted from a quantum theory, from the strongly excited molecules inside metallic nanocavities.

Metallic nanocavities are usually formed by nanoparticle dimers~\citep{HXu,WZhu,HHJeong}, nanoparticles-on-mirror (NPoM) constructs~\citep{NKongsuwan,JBLassiter} and STM tip-substrate structures~\citep{XWang,ZZhang}. These structures receive considerable attentions in recent years because they can enhance the electromagnetic fields by hundred folds, and concentrate them in tens of manometers. The huge field enhancement has been explored in surface-enhanced Raman~\citep{Moskovits1985}, and fluorescence spectroscopy ~\citep{EFort}, while the strong field concentration has been applied to enhance the light-matter interaction to demonstrate vacuum Rabi splittings \citep{SSavasta,AESchlather,RChikkaraddy,OSOjambati}, and the molecular optomechanical effects~\citep{FBenz,XLiu,YXu}.

In contrast to the experiments~\citep{YLuo}, we consider  the ultrafast plasmon-mediated superradiant pulses from the vertically standing methylene blue (MB) molecules inside a nanoparticle-on-mirror nanocavity (Fig. \ref{fig:system}), where such molecular configuration can be realized e.g. by capsulating the molecules in cucurbit[7]uril cages~\citep{RChikkaraddy}. To demonstrate our proposal, we develop a quantum theory for the system by combining the macroscopic quantum electrodynamics theory~\citep{SScheel,NRivera} and the open quantum system theory~\citep{HBreuer}, and we equip the theory with the numerical electromagnetic simulations of the realistic nanocavities via the boundary element methods~\citep{FGarcia,JWaxenegger}. 

Our calculation shows that in such a system, the molecules  experience a plasmonic Lamb shift by about $15$ meV, and a Purcell-enhanced decay rate by about $7$ meV, and can interact with each other via coherent and dissipative coupling, which are mediated by either the surface plasmons or the free-space field.  More importantly, the coherent coupling mediated by short-ranged propagating surface plasmons cancels largely the typical dipole-dipole coupling mediated by the free-space field, strongly suppressing the formation of delocalized excitons, and the dominated dissipative coupling mediated by the relatively long-ranged gap plamsons enables the ultra-fast superradiant pulses within pico-second scale from the strongly excited molecules. Thus, our study opens up the possibility of studying the rich
superradiant effects from the quantum emitters in a sub-wavelength scale by engineering the plasmonic environments. 


\paragraph{Quantum Master Equation.---}

In the Appendix \ref{sec:App1}, we achieve an effective master equation for   the reduced density operator $\hat{\rho}$ of the molecules by eliminating adiabatically the electromagnetic field
reservoir of the metallic nanocavity. Incorporating the excitation of the molecules in a semi-classical way, we extend this equation as: $ \frac{\partial}{\partial t}\hat{\rho}  =-\frac{i}{\hbar}\left[\hat{H}_{mol}+\hat{H}_{las}+\hat{H}_{pla},\hat{\rho}\right] +\mathcal{D}_{plas}\left[\hat{\rho}\right]. $ The Hamiltonian $\hat{H}_{mol}=\sum_{s=1}^{N} \hbar\omega_{s} \hat{\sigma}_{s}^{22}$ describes the $N$ molecules (labeled by $s$) with frequency $\omega_{s}$ and projection operator $\hat{\sigma}_{s}^{22}$, where the ground and excited state are labeled by the upper indices $1,2$. For the sake of simplicity, we assume that all the molecules have the same transition frequency $\omega_s=\omega_e$. The  Hamiltonian
$\hat{H}_{las}=\hbar\sum_{s}\left(\hat{\sigma}_{s}^{21}v_{s}e^{-i\omega_{l}t}+v_{s}^{*}e^{i\omega_{l}t}\hat{\sigma}_{s}^{12}\right)$ describes the optical excitation of the molecules with the raising $\hat{\sigma}_{s}^{21}$ and lowering operator $\hat{\sigma}_{s}^{12}$, and the coefficients  $\hbar v_{s}=-\mathbf{d}_{s}\cdot\mathbf{E}\left(\mathbf{r}_{s},\omega_{l}\right)$, which are determined by the molecular transition dipole moment $\mathbf{d}_{s}$ and the  enhanced (classical) electric field at the molecular position $\mathbf{r}_{s}$ (excited by a laser of frequency $\omega_{l}$).

The Hamiltonian $\hat{H}_{pla}=- \sum_{s,s'=1}^{N}\hbar\Omega_{ss'}\hat{\sigma}_{s}^{21}\hat{\sigma}_{s'}^{12}$ accounts for the reduction of molecular excitation energy by $\Omega_{ss}$, i.e. plasmonic Lamb shift~\citep{YZhang2017}, and the inter-molecular coherent coupling $\Omega_{ss'}$ ($s\neq s'$). The dissipative term $\mathcal{D}_{plas}\left[\hat{\rho}\right]=\sum_{s,s'=1}^{N}\frac{1}{2}\Gamma_{ss'}\left(2\hat{\sigma}_{s'}^{12}\hat{\rho}\hat{\sigma}_{s}^{21}-\hat{\sigma}_{s}^{21}\hat{\sigma}_{s'}^{12}\hat{\rho}-\hat{\rho}\hat{\sigma}_{s}^{21}\hat{\sigma}_{s'}^{12}\right)$ describes the Purcell-enhanced molecular decay rate $\Gamma_{ss}$~\citep{BYang}, and the inter-molecular dissipative coupling $\Gamma_{ss'}$ ($s'\neq s$). All these processes are pictorially illustrated in Fig.~\ref{fig:system}.  $
\Omega_{ss'} =\frac{1}{\hbar\epsilon_{0}}\left(\frac{\omega_{e}}{c}\right)^{2}\mathbf{p}_{s}^{*}\cdot\mathrm{Re} \overleftrightarrow{G}\left(\mathbf{r}_{s},\mathbf{r}_{s'};\omega_{e}\right)\cdot\mathbf{p}_{s'}$ is determined by the vacuum permittivity $\epsilon_{0}$, the light speed $c$, and the dyadic Green's tensor $\overleftrightarrow{G}\left(\mathbf{r}_{s},\mathbf{r}_{s'};\omega_{e}\right)$. The parameters $\Gamma_{ss'}/2$ follow similar expression except for taking the imaginary part of Green's tensor $\mathrm{Im} \overleftrightarrow{G}\left(\mathbf{r}_{s},\mathbf{r}_{s'};\omega_{e}\right)$. For the sake of simplicity, we have ignored the intrinsic decay and dephasing of molecules in our study. 

As explained in the Appendix \ref{sec:spectrum}, the far-field radiation (detected at the position $\mathbf{r} $)
$I(t) \approx \mathrm{Re}\sum_{s,s'=1}^{N}K_{ss'}\left\langle \hat{\sigma}_{s'}^{21}\hat{\sigma}_{s}^{12}\right\rangle
(t)$ can be calculated with the propagation factor 
$ K_{ss'}  =\frac{cr^{2}}{4\pi^{2}\epsilon_{0}}\left[\frac{\omega_{s'}^{2}}{c^{2}}\overleftrightarrow{G}^{*}\left(\mathbf{r},\mathbf{r}_{s'};\omega_{s'}\right)\cdot\mathbf{d}_{s'}\right] \cdot\left[\frac{\omega_{s}^{2}}{c^{2}}\overleftrightarrow{G}\left(\mathbf{r},\mathbf{r}_{s};\omega_{s}\right)\cdot\mathbf{d}_{s}^{*}\right]$, and the expectation values $\left\langle \hat{\sigma}_{s'}^{21}\hat{\sigma}_{s}^{12}\right\rangle(t) = {\rm tr}\left\{ \hat{\sigma}_{s'}^{21}\hat{\sigma}_{s}^{12} \hat{\rho} (t)\right\}$. To clarify the collective nature of radiation, we split the total radiation into the contribution from the individual molecules  $I_{ind} \approx \mathrm{Re}\sum_{s}^{N}K_{ss}\left\langle \hat{\sigma}_{s}^{21}\hat{\sigma}_{s}^{12}\right\rangle(t)$ and the interference of molecules  $I_{int} \approx \mathrm{Re}\sum_{s \neq s'}^{N}K_{ss'}\left\langle \hat{\sigma}_{s'}^{21}\hat{\sigma}_{s}^{12}\right\rangle
(t)$, which are determined by the diagonal terms with $s=s'$, and the off-diagonal terms with $s \neq s'$, respectively. To simulate as much molecules as possible, we solve the quantum master equation with the mean-field approach~\citep{DPl} instead of the standard density matrix method, see the Appendix \ref{sec:codes}.

\begin{figure}[!htp]
\centering
\includegraphics[scale=0.305]{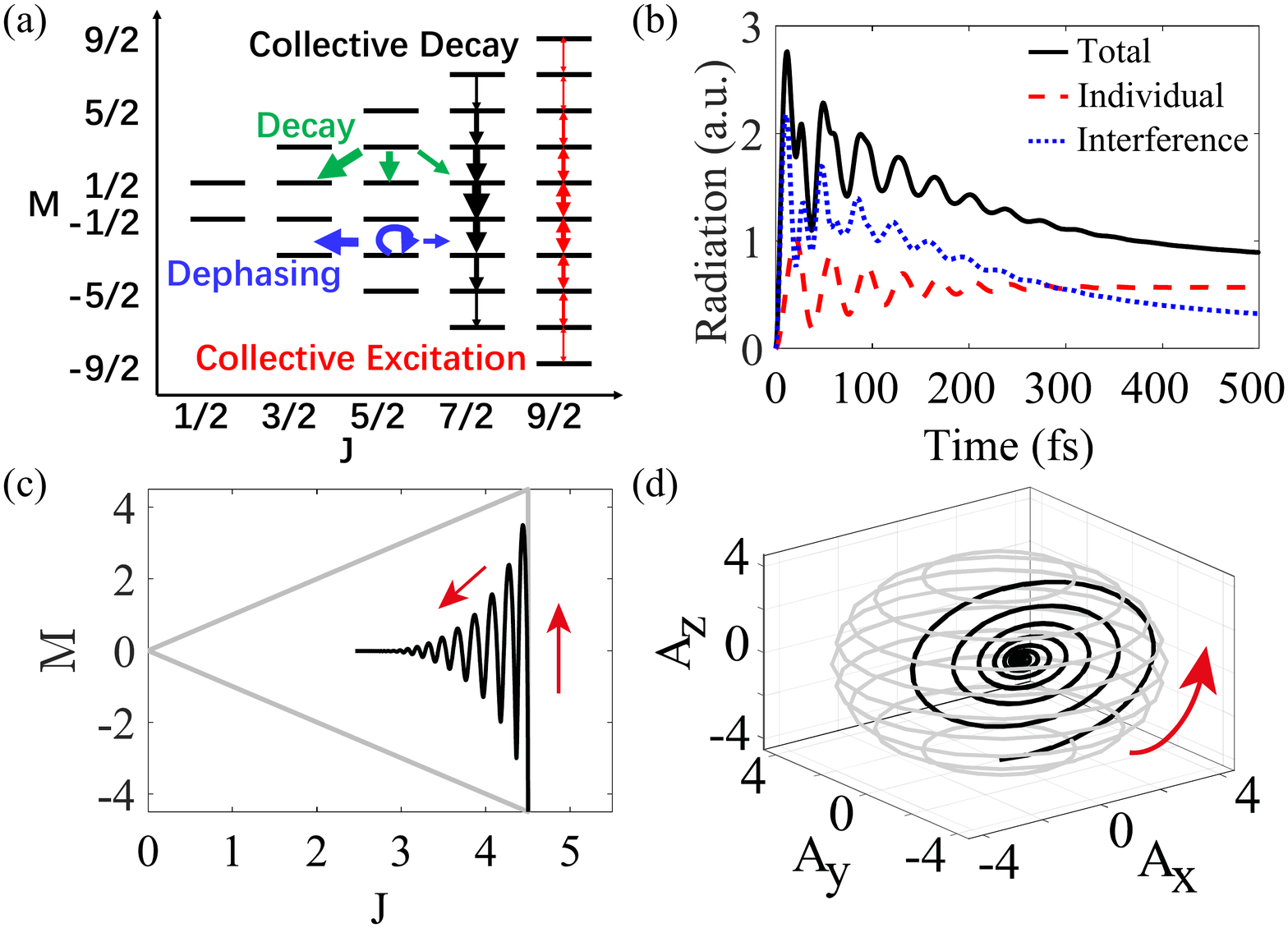}
\caption{\label{fig:continuous} Continuous superradiance for a square array of nine molecules with $1$ nm separation in the nanocavity center. Panel (a) shows the Dicke states for an ideal system with nine identical molecules,  and the transitions due to the collective excitation (red arrows), the quantum jumps due to the collective decay (black arrows), and the individual decay (green arrows) as well as the individual dephasing (blue arrows). Here, the  thickness of arrows shows the relative strength, and the mentioned processes can occur for all the Dicke states with the strengths weighted by the factors given in Ref.~\citep{KDebnath}. Panel (b) shows the dynamics of the far-field radiation (black solid line), and the contributions of the molecular interference (blue dotted line) and the individual molecules (red dashed line). Panel (c) and (d) show the dynamics of the molecules with  the average of Dicke state quantum numbers and the collective spin vector, where the gray lines indicate the out-most sphere and the Dicke states boundaries, respectively. Here, the laser illumination with $10^4$ $\mu W/\mu m^2$ is assumed to be resonant with the BQP mode. }
\end{figure}

To illustrate the collective dynamics of the molecular ensemble, we introduce the collective spin operators $\hat{J}_{x}=(1/2)\sum_{s}\left(\hat{\sigma}_{s}^{12}+\hat{\sigma}_{s}^{21}\right)$, $\hat{J}_{y}=(i/2)\sum_{s}\left(\hat{\sigma}_{s}^{12}-\hat{\sigma}_{s}^{21}\right)$ and $\hat{J}_{z}=(1/2)\sum_{s}\left(2\hat{\sigma}_{s}^{22}-1\right)$, and calculate the collective spin vector ${\bf A}=\sum_{i=x,y,z}\langle\hat{J}_{i}\rangle {\bf e}_{i}$ with their expectation values $\langle\hat{J}_{i}\rangle$ and the unit vectors ${\bf e}_{i}$ in the Cartesian coordinate system. The collective spin vector traces a spherical surface with the center at the origin and the radius $N/2$ for the molecules in the pure quantum states. It points to the south and north pole of the sphere for the molecules on the fully ground and excited state, respectively, and points to the positions inside the sphere for the molecules on the mixed states or entangled states [see Fig.~\ref{fig:continuous}(d)]. In addition, we employ also the Dicke states~\citep{RHDicke1954} $\left|J,M \right\rangle$ to interpret the dynamics of the molecular ensemble, where the integer and half-integer $J,M$ in the range $J \le N/2$ and $-J<M<J$ describe the symmetry and the excitation degree of the Dicke states. Usually, the states for given $J$ and different $M$ are illustrated as a ladder with equal spacing, and those for different $J$ as shifted ladders,  all forming a triangle space [see Fig.~\ref{fig:continuous}(a)]. In the following, we  calculate the average of the Dicke state quantum numbers with the relations $\overline{J}(\overline{J}+1) = \sum_{i}\langle\hat{J}_{i}^2\rangle$, $\overline{M}=\langle\hat{J}_{z}\rangle$, see the Appendix~\ref{sec:meanequations} for the exact expressions. 


\begin{figure}[!htp]
\begin{centering}
\includegraphics[scale=0.30]{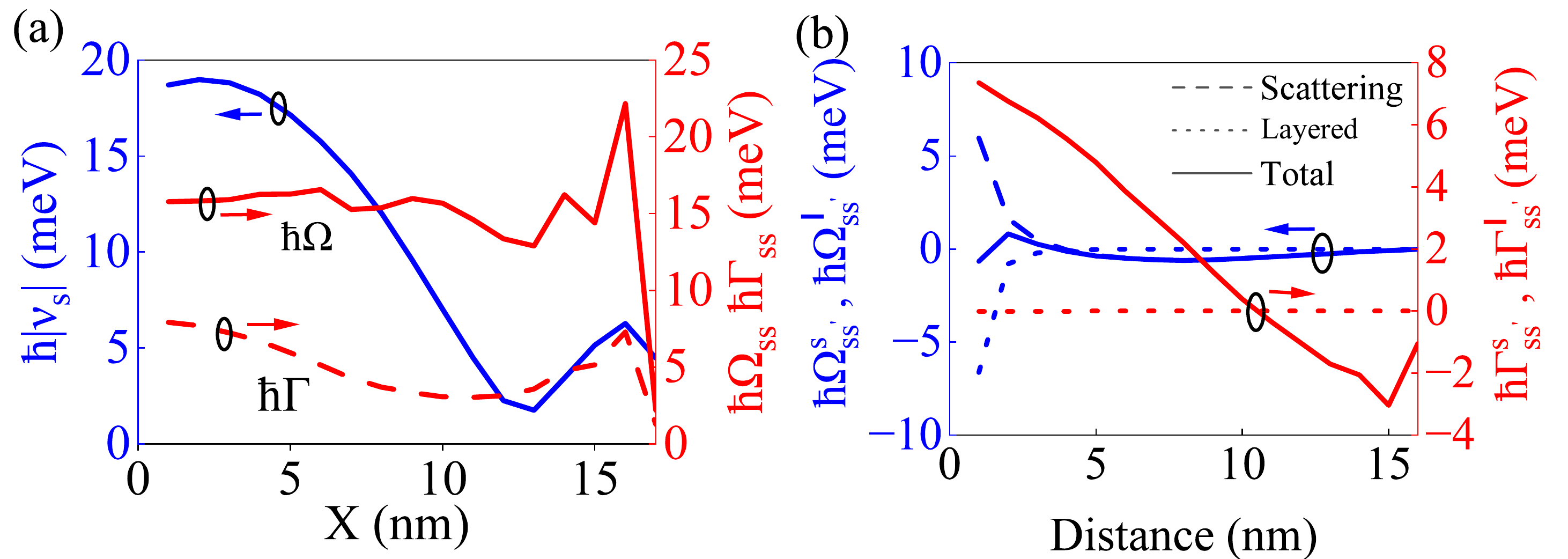}
\par\end{centering}
\caption{ \label{fig:params} Molecular parameters. Panel (a) shows the excitation coefficient $\hbar v_s$ of one molecule moving along the x-axis, for the laser at BQP wavelength and intensity $I_{las}= 10^4 \mu W/\mu m^2$  (blue line), and the corresponding plasmonic shift  $\Omega_{ss}$ and Purcell-enhanced decay rate  $\Gamma_{ss}$ (red solid and dashed line).  Panel (b) shows the coherent coupling  $\Omega_{ss'}$ (blue lines) and  dissipative coupling   $\Gamma_{ss'}$ (red lines), mediated by the scattered field (with the upper index "$s$", solid lines) and the field associated with the layered structure (with the upper index "l", dashed lines), for one molecule at the center and another molecule moving away along the x-axis.  All the molecules have the transition dipole of $4$ Debye. For more details, see the text. }
\end{figure}

\paragraph{Excitation Coefficient, Coherent and Dissipative Coupling of Molecules.---}

We utilize the  boundary element method (BEM)~\citep{FJGDAbajo,FJGDAbajo1}, as implemented in the metal nanoparticle BEM toolkit ~\citep{UHohenester,JWaxenegger}, and together with the  dielectric permittivity of gold as determined in the experiment~\citep{PBJohnson}, and of $2.1$ for the nanogap, to carry out the electromagnetic simulations  for the NPoM nanocavity shown in~Fig. \ref{fig:system}. In the Appendix~\ref{sec:plasmon}, we provide a detailed analysis of the plasmonic response, and summarize shortly in the following. We find that the electric field along the vertical direction in the nanocavity is dominated by the bonding dipole plasmon (BDP) and the bonding quadruple plasmon (BQP)~\citep{FBenz} at wavelength $820$ nm, $660$ nm, respectively. The dyadic Green's tensor can be decomposed to the contributions of the field scattered off the truncated nanosphere, and the field associated with the layered structure~\citep{JWaxenegger}, which can further be split into those due to the free-space field and the field reflected off the interfaces of the layered structure~\citep{MPaulus}. For the vertical tensor component at the middle of nanocavity, the former with same positions  shows peaks and Fano features in the real and imaginary part at the BDP and BQP wavelength over a smooth background, which can be attributed to the propagating surface plasmon of the corresponding layered structure~\citep{YZhang2021}. In contrast, the latter with with same positions diverges due to the free-space field contribution, and the one with two slightly different positions shows negative real part without obvious wavelength-dependence, which can again be attributed to the free-space field, and negative imaginary part with feature below $600$ nm, which is due to the field reflected from the interfaces of layered structure. In addition, we have also examined the spatial dependence of the field enhancement and the Green's tensor for the BQP and BDP modes.

From the above results, we obtain the spatial dependence of the molecular excitation coefficient, the inter-molecular coherent and dissipative coupling [Fig.~\ref{fig:params}]. Here, we assume that the molecules are resonant to the BQP mode, and have a  transition dipole moment of $4$ Deybe~\citep{TBDQueiroz}. We find that the excitation coefficient $\hbar|\nu|$ follows the spatial distribution of the field enhancement [blue solid lines in Fig.~\ref{fig:params}(a)], and could amount to $60$ meV for the laser intensity $I_{las} = 10^{4}$ $\mu W/\mu m^2$, which is achievable in the experiments~\citep{NLombardi}. The plasmonic Lamb shift $\Omega_{ss}$ does not show obvious spatial dependence, while the Purcell-enhanced decay rate $\Gamma_{ss}$ shows the clear BQP pattern [blue solid and dashed lines in Fig.~\ref{fig:params}(a)]. In the evaluation of these parameters, we have considered only the contribution from the scattered field. Usually, the free-space field contribution of the layered Green's tensor diverges for the same positions, and this divergence can be regularized, leading to the well-known Lamb shift~\citep{PDeVries}. However, in the normal treatment, one can also consider that this shift is already included in the definition of the molecular transition frequency. 

Furthermore, we analyze the coherent and dissipative coupling [Fig.~\ref{fig:params}(b)]. We see that the free-space field-mediated coherent coupling is always negative, and reduces dramatically to zero for the inter-molecular distance within $4$ nm (blue dotted line), while the scattered field-mediated coherent coupling is positive and decays also dramatically for the short inter-molecular distance (blue dashed line). As a result, the total coherent coupling is actually below $2$ meV for the molecules in such short distance. Note that for much larger inter-molecular distance, the coherent coupling is dominated by the gap plasmon, and is much smaller than $1$ meV. In contrast, the free-space field-mediated dissipative coupling is negligible (red dotted line), while the scattered field-mediated one is below $8$ meV (red dashed line). In addition, the dissipative coupling decays rather slowly with increasing inter-molecular distance. From these results, we might conclude that the dissipative coupling is long-ranged, and is comparable or larger than the coherent coupling. In contrast, for the molecules in the free-space the coherent coupling will dominate.

\paragraph{Continuous Superradiance.---}

After obtaining the related parameters, we solve now the second-order mean-field equations to analyze the system response to the laser excitation. Since the simulations with more than ten molecules are time consuming, here, we consider a square array of nine molecules with $1$ nm separation in the middle of the nanocavity. We have checked that the superradiance as explained blow is not captured by the first-order mean-field equations (not shown). 

Before presenting the results, it is worth of recapturing what was known on the superradiance of the ideal system with identical emitters, as revealed in our earlier work~\citep{KDebnath}. As shown in Fig.~\ref{fig:continuous}(a), the driving of the emitters by  an external field with same strength leads to the transitions between the Dicke states with higher strengths to the states in the middle of the states ladders, and the collective decay of the emitters leads to the quantum jumps between the Dicke states with higher probability for these states. In addition, the decay (dephasing) of individual emitters with identical rate leads to the quantum jumps to the Dicke states with reduced (unchanged) $M$ and $J$, which does not change or changes by one,  and the quantum jumps to the state with reduced $J$ has higher probability.

Under the continuous laser excitation with the intensity $10^{4}$ $\mu W/\mu m^2$, we find with Fig.~\ref{fig:continuous}(b) that the total radiation shows complex oscillations and smooth decay for short and long time (solid line), and the contribution of molecular interference is about twice larger than that of individual molecules (dotted vs dashed line). In the Dicke states space, the molecules move almost vertically upwards from the lower-right corner (ground state) along the right boundary (i.e. so-called superradiant states), and then moves almost vertically downwards to a point slightly higher than the lower boundary, and finally repeat the same vertical dynamics until reaching the inner Dicke states with $M\approx0$ [Fig.~\ref{fig:continuous} (c)]. In comparison to Fig.~\ref{fig:continuous} (a), this dynamics can be understood as a consequence of the competence of the aforementioned four processes, where the collective excitation and decay originate from the average coupling with the gap plasmon,  the individual decay can be attributed to the deviation from the average coupling for individual molecules, the dephasing can be interpreted by the slightly different plasmonic Lamb shift for the individual molecules. At the same time, the collective spin vector rotates around an axis in the x-y plane, which orientate roughly $45$ degree with respect to the x- and y-axis, and the rotation starts from the south pole (fully ground state) on the out-most spherical surface, and then on the inner spherical surfaces with reduced radius  [Fig.~\ref{fig:continuous} (d)].

\begin{figure}[!htp]
\centering
\includegraphics[scale=0.30]{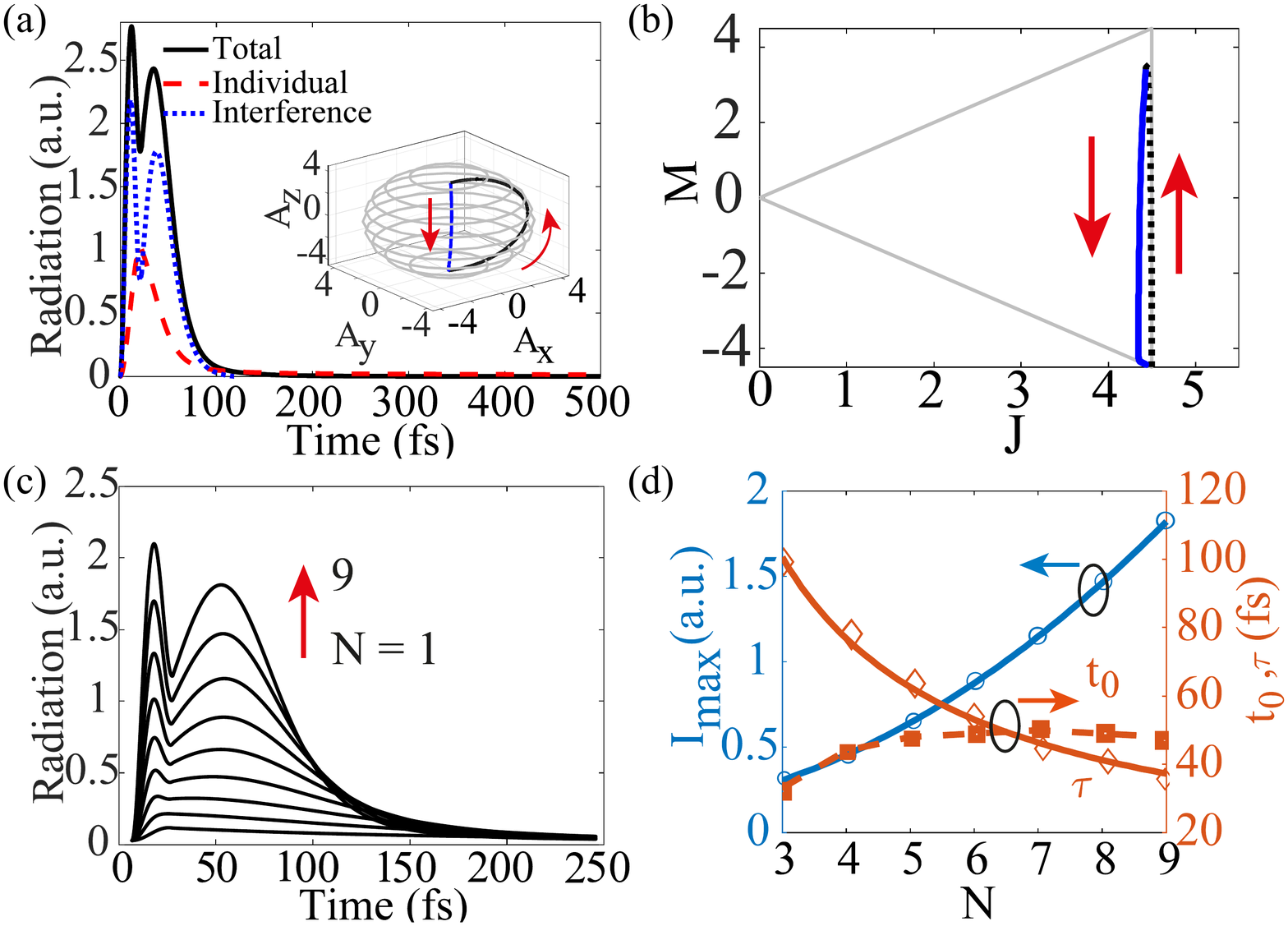}
\caption{\label{fig:pulses} Pulsed superradiance for a square array of nine molecules with one nanometer  separation in the nanocavity center. Panel (a-c) show the similar results as Fig.~\ref{fig:continuous}(b-d) except for  a laser pulse excitation of $22$ fs duration. Panel (c) shows the radiation for system with increasing number of molecules $N$ (lower to upper curves). Panel (d) shows the pulse maximum $I_{max}$ (blue circles), the pulse center $t_0$ and duration $\tau$ (red diamonds and green squares) as function of $N$, where the solid lines are fitting curves. Here, the laser illumination with $10^4$ $\mu W/\mu m^2$ is assumed to be resonant with the BQP mode.
}
\end{figure}

\paragraph{Ultrafast Superradiant Pulses.---}

From the dynamics revealed above, we notice that the molecules can be excited to reach as close as possible to the fully excited states. This affords us the possibility of studying the superradiant pulses from the strongly excited molecules. To this end, we drive the molecules with a laser pulse of $22$ fs long, and then analyze the dynamics thereafter.  We observe a superradiant pulse with center around $50$ fs and width of about $38$ fs, which mainly comes from the molecular interference [Fig.~\ref{fig:pulses} (a)].  To analyze the cause of this pulse, we further analyze the dynamics of molecules, and find that the collective spin vector withdraws firstly and stretches  along the z-axis [inset of Fig.~\ref{fig:pulses} (a)], and the molecular ensemble moves almost vertically downwards in the Dicke state space [Fig.~\ref{fig:pulses} (b)], which is caused by the collective decay of the molecules [Fig.~\ref{fig:continuous} (a)].

We  examine  further  the dependence of the superradiant pulses on the number of molecules $N$. With increasing $N$, the superradiant pulses become stronger and slightly narrower, and appear also slightly latter [Fig.~\ref{fig:pulses} (c)]. To quantify these changes, we have extracted the pulse maximum, center and width, and plotted them as function of $N$ [Fig.~\ref{fig:pulses} (d)]. Note that for $N \le 3$, the pulse is not obvious, and thus is excluded. We see that the pulse maximum increases super-linearly, and the pulse width reduces steadily, while the pulse center increases firstly and then decays slightly. The detailed analysis indicates that the pulse maximum and width follow the expressions $\sim 0.12 + 0.02N^2$, and $5.59 +  285.10/N$, respectively. The observed quadratic and inverse scaling are expected for the superradiant pulses~\citep{AVAndreev1980}, and are also consistent with other experiments~\citep{MANorcia2016}. We have also considered the situation that the molecules are resonant to the BDP mode, and found the similar results (Fig.~\ref{fig:sup-BDP}), except that the obtained superradiant pulses are about three times longer due to the relatively small dissipative coupling between the molecules. Thus, by searching the metallic nanostructures leading to much smaller dissipative coupling, the superradiant pulses might be further elongated to e.g. picosecond scale. In addition, we have also examined a linear array of molecules in the middle of nanocavity, and found that the spatial dependence of the excitation coefficient has strong influence on the molecular response and the resulting superradiant signal (Fig.~\ref{fig:sup-linear}). 

\paragraph{Conclusions.---}

In summary, we demonstrated theoretically the ultrafast plasmon-mediated superradiant pulses from vertically standing molecules inside a metallic nanocavity. In this system, the short-ranged guiding plasmon-mediated coherent coupling between the molecules cancels largely the free-space field-mediated one, and  the long-ranged gap plasmon-mediated dissipative coupling dominates and establishes the quantum correlations between the molecular pairs during the collective decay, leading to the ultra-fast superradiant pulses.  

On the basis of the current work, several interesting  phenomena  might be explored in future. Firstly, one might introduce the incoherent molecular pumping to compensate the collective decay to form the steady-state superradiance~\citep{DMeiser,JGBohnet}. Second, one can explore the interplay between the superradiant effects and the strong molecules-plasmon coupling, where the cancellation of the coherent couplings might play also an important role. Finally, by incorporating the electron-vibration coupling, one can also explore the influence of the superradiant effects on the plasomn-enhanced resonant Raman scattering.

Yuan Zhang convinced the idea, the theory and the program, YuXin Niu calculated all the results. All the authors contribute to the writing of the manuscript. We acknowledge the fruitful discussions with Dr. Ruben Esteban, and the financial support by the National Key R\&D Program of China under Grant No. 2021YFA1400900, the National Natural Science Foundation of China through the project No. 12004344, 21902148, 12074232, 12125406, and the NSFC-DPG joint project No. 21961132023.

\appendix
\renewcommand\thefigure{A\arabic{figure}}
\renewcommand\thetable{A\arabic{table}}
\renewcommand{\thepage}{A\arabic{page}}
\renewcommand{\bibnumfmt}[1]{[S#1]}
\setcounter{figure}{0}  

\section{Molecular Master Equation by Adiabatically Eliminating Field Reservoir
\label{sec:App1}}

In this section, we present the derivation of the superradiant master
equation for many molecules inside a NPoM plasmonic nano-cavity. According to the macroscopic quantum electrodynamics theory~\citep{SScheel,NRivera}, the electromagnetic field can be described as a continuum via the Hamiltonian 
\begin{equation}
\hat{H}_{f}=\int d\mathbf{r}\int_{0}^{\infty}d\omega_{f}\hbar\omega_{f}\hat{\mathbf{f}}^{\dagger}\left(\mathbf{r},\omega_{f}\right)\cdot\hat{\mathbf{f}}\left(\mathbf{r},\omega_{f}\right)
\end{equation}
with frequency $\omega_{f}$, creation $\hat{\mathbf{f}}^{\dagger}\left(\mathbf{r},\omega_{f}\right)$
and annihilation $\hat{\mathbf{f}}\left(\mathbf{r},\omega_{f}\right)$
noise (bosonic) operators with at position $\mathbf{r}$, and the
quantized electric field operator 
\begin{align}
 & \hat{\mathbf{E}}\left(\mathbf{r},\omega_{f}\right)=i\sqrt{\frac{\hbar}{\pi\epsilon_{0}}}\frac{\omega_{f}^{2}}{c^{2}}\int d\mathbf{r}'\sqrt{{\rm Im} \epsilon\left(\mathbf{r}',\omega_{f}\right)}\nonumber \\
 & \times\overleftrightarrow{G}\left(\mathbf{r},\mathbf{r}';\omega_{f}\right)\cdot\hat{\mathbf{f}}\left(\mathbf{r}',\omega_{f}\right),\label{eq:Eoperator}
\end{align}
is determined by the imaginary part of the dielectric function ${\rm Im} \epsilon\left(\mathbf{r}',\omega\right)$
, the classical dyadic Green's function $\overleftrightarrow{G}\left(\mathbf{r},\mathbf{r}';\omega\right)$. 

To study the interaction of two molecules with the NPoM nano-cavity,
we model the molecules as two-level systems via the Hamiltonian $\hat{H}_{mol}=\sum_{s=1}^{N}\left(\hbar\omega_{s}\right)\hat{\sigma}_{s}^{22}$, where the frequency $\omega_{s}$ and the projection operator $\hat{\sigma}_{22}^{z}$ are associated with the $s$-th molecule. In the rotating wave approximation, the molecules interact with the quantized field via the Hamiltonian
\begin{align}
\hat{H}_{fm} & =-\sum_{s=1}^{2}\int_{0}^{\infty}d\omega_{f}\bigl[\hat{\sigma}_{s}^{21}\mathbf{d}_{s}\cdot\hat{\mathbf{E}}\left(\mathbf{r}_{s},\omega_{f}\right)\nonumber \\
 & +\mathbf{d}_{s}^{*}\cdot\hat{\mathbf{E}}^{\dagger}\left(\mathbf{r}_{s},\omega_{f}\right)\hat{\sigma}_{s}^{12}\bigr],
\end{align}
where $\hat{\sigma}_{s}^{12},\hat{\sigma}_{s}^{21},\mathbf{d}_{s}$ are the lowing and raising operator as well as the transition dipole moment of the $s$-th molecule. 

To reduce the degree of freedom under consideration, we will treat
the electromagnetic field as reservoir and obtain an effective master
equation for the molecules by adiabatic-ally eliminating the reservoir
degree of freedom. To this end, we firstly consider the Heisenberg
equation for the operator $\hat{O}$ of the molecules 
\begin{align}
 & \frac{\partial}{\partial t}\hat{O}\left(t\right)=\left[\sum_{s}\hat{\sigma}_{s}^{21}\left(t\right),\hat{O}\left(t\right)\right]\int_{0}^{\infty}d\omega_{f}\frac{\omega_{f}^{2}}{c^{2}}\int d\mathbf{r}'\nonumber \\
 & \times\sqrt{\frac{{\rm Im} \epsilon\left(\mathbf{r}',\omega_{f}\right)}{\hbar\pi\epsilon_{0}}}\mathbf{d}_{s}\cdot\overleftrightarrow{G}\left(\mathbf{r}_{s},\mathbf{r}';\omega_{f}\right)\cdot\hat{\mathbf{f}}\left(\mathbf{r}',\omega_{f},t\right)\nonumber \\
 & -\sum_{s=1}^{2}\mathbf{d}_{s}^{*}\cdot\int_{0}^{\infty}d\omega_{f}\frac{\omega_{f}^{2}}{c^{2}}\int d\mathbf{r}'\sqrt{\frac{{\rm Im} \epsilon\left(\mathbf{r}',\omega_{f}\right)}{\hbar\pi\epsilon_{0}}}\nonumber \\
 & \overleftrightarrow{G}^{*}\cdot\left(\mathbf{r},\mathbf{r}';\omega_{f}\right)\hat{\mathbf{f}}^{\dagger}\left(\mathbf{r}',\omega_{f},t\right)\left[\hat{\sigma}_{s}^{12}\left(t\right),\hat{O}\left(t\right)\right].\label{eq:eq-o}
\end{align}
This equation depends on the field operators $\hat{\mathbf{f}}\left(\mathbf{r}',\omega_{f},t\right)$
[and its conjugation
$\hat{\mathbf{f}}^{\dagger}\left(\mathbf{r}',\omega_{f},t\right)$],
which follows the following Heisenberg equation 
\begin{align}
 & \frac{\partial}{\partial t}\hat{\mathbf{f}}\left(\mathbf{r},\omega_{f},t\right)=-i\omega_{f}\hat{\mathbf{f}}\left(\mathbf{r},\omega_{f},t\right)\nonumber \\
 & +\sum_{s=1}^{2}\frac{\omega_{f}^{2}}{c^{2}}\sqrt{\frac{{\rm Im} \epsilon\left(\mathbf{r},\omega_{f}\right)}{\hbar\pi\epsilon_{0}}}\mathbf{d}_{s}^{*}\cdot\overleftrightarrow{G}^{*}\left(\mathbf{r}_{s},\mathbf{r};\omega_{f}\right)\hat{\sigma}_{s}^{12}\left(t\right).\label{eq:eqn-f}
\end{align}
where we have used the commutation relations  
\begin{align}
\left[\hat{\mathbf{f}}\left(\mathbf{r}',\omega_{f}',t\right),\hat{\mathbf{f}}\left(\mathbf{r},\omega_{f},t\right)\right] & =0,\\
\left[\hat{\mathbf{f}}^{\dagger}\left(\mathbf{r}',\omega'_{f},t\right),\hat{\mathbf{f}}\left(\mathbf{r},\omega_{f},t\right)\right] & =-\delta\left(\mathbf{r}-\mathbf{r}'\right)\delta\left(\omega_{f}-\omega_{f}'\right).
\end{align}
 The formal solution of Eq. (\ref{eq:eqn-f}) is 

\begin{align}
 & \hat{\mathbf{f}}\left(\mathbf{r},\omega_{f},t\right)=\sum_{s=1}^{2}\frac{\omega_{f}^{2}}{c^{2}}\sqrt{\frac{{\rm Im} \epsilon\left(\mathbf{r},\omega_{f}\right)}{\hbar\pi\epsilon_{0}}}\mathbf{d}_{s}^{*}\cdot\overleftrightarrow{G}^{*}\left(\mathbf{r}_{s},\mathbf{r};\omega_{f}\right)\nonumber \\
 & \times\int_{0}^{t}dt'e^{-i\omega_{f}\left(t-t'\right)}\hat{\sigma}_{s}^{12}\left(t'\right),\label{eq:feq-f}
\end{align}
The equation for the conjugate field operator $\hat{\mathbf{f}}^{\dagger}\left(\mathbf{r},\omega_{f},t\right)$
and its formal solution can be achieved by taking the conjugation
over Eq. (\ref{eq:eqn-f}) and (\ref{eq:feq-f}). 

At this moment, if we insert Eq. (\ref{eq:feq-f}) into Eq. (\ref{eq:eq-o}),
we will obtain a differential and integral equation. By solving this
equation, we are able to study not only Markov dynamics in the weak
coupling regime, but also the non-Markov dynamics in the strong coupling
regime. Since here we focus on the former regime, we carry out the
Born-Markov approximation to the formal solution (\ref{eq:feq-f}).
To do so, we replace $\hat{\sigma}_{s}^{12}\left(\tau\right)$ by $e^{\omega_{s}\left(t-\tau\right)}\hat{\sigma}_{s}^{12}\left(t\right)$
in this expression, and then define a new variable $\tau=t-t'$ to
change the integration over the time, and finally change the upper
limit of this integration into infinity to achieve the following expression
\begin{align}
 & \hat{\mathbf{f}}\left(\mathbf{r},\omega_{f},t\right)\approx\sum_{s=1}^{2}\frac{\omega_{f}^{2}}{c^{2}}\sqrt{\frac{{\rm Im} \epsilon\left(\mathbf{r},\omega_{f}\right)}{\hbar\pi\epsilon_{0}}}\mathbf{d}_{s}^{*}\cdot\overleftrightarrow{G}^{*}\left(\mathbf{r}_{s},\mathbf{r};\omega_{f}\right)\nonumber \\
 & \times\hat{\sigma}_{s}^{12}\left(t\right)\left(\pi\delta\left(\omega_{s}-\omega_{f}\right)+i\mathcal{P}\frac{1}{\omega_{s}-\omega_{f}}\right).\label{eq:fBM}
\end{align}

In the last step, we have have utilized the relation 
\begin{equation}
\int_{0}^{\infty}d\tau e^{i\left(\omega_{s}-\omega_{f}\right)\tau}=\pi\delta\left(\omega_{s}-\omega_{f}\right)+i\mathcal{P}\frac{1}{\omega_{s}-\omega_{f}}.
\end{equation}
Inserting Eq. \eqref{eq:fBM} (and its conjugation) into Eq. (\ref{eq:eq-o}), using the property of the dyadic Green's function
\begin{align}
 & \left(\frac{\omega_{f}}{c}\right)^{2}\sum_{j}\int d^{3}\mathbf{r}'{\rm Im} \epsilon\left(\mathbf{r}',\omega_{f}\right)G_{k'j}\left(\mathbf{r}_{1},\mathbf{r}';\omega_{f}\right)\nonumber \\
 & G_{kj}^{*}\left(\mathbf{r}_{2},\mathbf{r}';\omega_{f}\right)=\mathrm{Im}G_{k'k}\left(\mathbf{r}_{1},\mathbf{r}_{2};\omega_{f}\right),\label{eq:identity}
\end{align}
and applying the Kramer-Kronig relation 
\begin{align}
 & \mathcal{P}\int d\omega_{f}\frac{d\omega_{f}}{\omega_{f}-\omega_{s'}}\frac{\omega_{f}^{2}}{c^{2}}\mathbf{d}_{s}\cdot\mathrm{Im}\overleftrightarrow{G}\left(\mathbf{r}_{s},\mathbf{r}_{s'};\omega\right)\cdot\mathbf{d}_{s'}^{*}\nonumber \\
 & =\pi\frac{\omega_{s'}^{2}}{c^{2}}\mathbf{d}_{s}\cdot\mathrm{Re}\overleftrightarrow{G}\left(\mathbf{r}_{s},\mathbf{r}_{s'};\omega_{s'}\right)\cdot\mathbf{d}_{s'}^{*},\label{eq:KKrelation}
\end{align}
we obtain the following effective mater equation 
\begin{align}
\frac{\partial}{\partial t}\hat{O}\left(t\right) & =-i\sum_{s,s'=1}^{N}\left[\hat{\sigma}_{s}^{21}\left(t\right),\hat{O}\left(t\right)\right]\hat{\sigma}_{s'}^{12}\left(t\right)J_{ss'}^{\left(1\right)}\left(\omega_{s'}\right)\nonumber \\
 & -i\sum_{s,s'=1}^{N}J_{s's}^{\left(2\right)}\left(\omega_{s'}\right)\hat{\sigma}_{s'}^{21}\left(\tau\right)\left[\hat{\sigma}_{s}^{12}\left(t\right),\hat{O}\left(t\right)\right],\label{eq:emq}
\end{align}
where we have introduced the spectral densities 
\begin{align}
J_{ss'}^{\left(1\right)}\left(\omega\right) & =\frac{1}{\hbar\epsilon_{0}}\frac{\omega^{2}}{c^{2}}\mathbf{d}_{s}\cdot\overleftrightarrow{G}\left(\mathbf{r}_{s},\mathbf{r}_{s'};\omega\right)\cdot\mathbf{d}_{s'}^{*},\\
J_{ss'}^{\left(2\right)}\left(\omega\right) & =\frac{1}{\hbar\epsilon_{0}}\frac{\omega^{2}}{c^{2}}\mathbf{d}_{s}\cdot\overleftrightarrow{G}^{*}\left(\mathbf{r}_{s},\mathbf{r}_{s'};\omega\right)\cdot\mathbf{d}_{s'}^{*}.
\end{align}

In the next step, we consider the equation for the expectation value
$\mathrm{tr}\left\{ \hat{O}\left(t\right)\hat{\rho}\right\} =\mathrm{tr}\left\{ \hat{O}\hat{\rho}\left(t\right)\right\} $,
which can be computed either with the time-dependent operator $\hat{O}\left(t\right)$
and the time-independent density operator $\hat{\rho}$ in the Heisenberg
picture, or with the time-independent operator $\hat{O}$ and the time-dependent density operator in the Schrodinger picture. Using this relation and the cyclic property of the trace, we obtain the equation for the reduced density operator 
\begin{align}
\frac{\partial}{\partial t}\hat{\rho} & =-i\sum_{s,s'=1}^{N}\left[\hat{\sigma}_{s'}^{12}\hat{\rho},\hat{\sigma}_{s}^{21}\right]J_{ss'}^{\left(1\right)}\left(\omega_{s'}\right)\nonumber \\
 & -i\sum_{s,s'=1}^{N}J_{ss'}^{\left(2\right)}\left(\omega_{s}\right)\left[\hat{\rho}\hat{\sigma}_{s}^{21},\hat{\sigma}_{s'}^{12}\right].\label{eq:meq}
\end{align}
Introducing the new parameters 
\begin{align}
\Omega_{ss'} & =[J_{ss'}^{\left(1\right)}\left(\omega_{s}\right)+J_{ss'}^{\left(2\right)}\left(\omega_{s'}\right)]/2,\\
\Gamma_{ss'} & =-i[J_{ss'}^{\left(1\right)}\left(\omega_{s}\right)-J_{ss'}^{\left(2\right)}\left(\omega_{s'}\right)],
\end{align}
we can rewrite the spectral densities as 
\begin{align}
J_{ss'}^{\left(1\right)}\left(\omega_{s}\right) & =\left(\Omega_{ss'}+i\Gamma_{ss'}/2\right),\\
J_{ss'}^{\left(2\right)}\left(\omega_{s'}\right) & = \left(\Omega_{ss'}-i\Gamma_{ss'}/2\right).
\end{align}
Inserting these expressions into Eq. (\ref{eq:meq}) , we achieve
the following effective master equation 
\begin{align}
\frac{\partial}{\partial t}\hat{\rho} & =-i\sum_{s=1}^{N}\omega_{s}\left[\hat{\sigma}_{s}^{21}\hat{\sigma}_{s}^{12},\hat{\rho}\right]+i\sum_{s,s'=1}^{2}\Omega_{ss'}\left[\hat{\sigma}_{s}^{21}\hat{\sigma}_{s'}^{12},\hat{\rho}\right]\nonumber \\
 & +\sum_{s,s'=1}^{N}\frac{1}{2}\Gamma_{ss'}\left(2\hat{\sigma}_{s'}^{12}\hat{\rho}\hat{\sigma}_{s}^{21}-\hat{\sigma}_{s}^{21}\hat{\sigma}_{s'}^{12}\hat{\rho}-\hat{\rho}\hat{\sigma}_{s}^{21}\hat{\sigma}_{s'}^{12}\right).\label{eq:master-equation}
\end{align}
By solving this equation, we are able to study the superradiant and
subradiant effect of many molecules in NPoM nano-cavity. 

\section{Far-field Spectrum \label{sec:spectrum}}

In this section, we present the derivation of the far-field radiation
from the molecules in the NPoM nano-cavities. According to {[}ref
to S. Hughes, and M O. Sculley{]}, the far-field spectrum can be computed
with 
\begin{equation}
\frac{dW}{d\Omega}\left(t\right)=\frac{c\epsilon_{0}r^{2}}{4\pi^{2}} \mathrm{tr}\left\{ \hat{\mathbf{E}}^{\dagger}\left(\mathbf{r},t\right)\cdot\hat{\mathbf{E}}\left(\mathbf{r},t\right)\hat{\rho}\right\} .\label{eq:spectrum}
\end{equation}
In this expression, $r$ is the distance between the molecules and
the detector, $\hat{\mathbf{E}}\left(\mathbf{r},\tau\right)=\int d\omega_{f}\hat{\mathbf{E}}\left(\mathbf{r},\omega_{f},\tau\right)$
is the electric field operator at the detector position, and is obtained
by the integration of the electric field over the frequency $\omega_{f}$.
Inserting Eq. (\ref{eq:fBM}) into Eq. (\ref{eq:Eoperator}), we obtain
the following expression

\begin{align}
 & \hat{\mathbf{E}}\left(\mathbf{r},\omega_{f},t\right)=i\frac{\hbar}{\pi\epsilon_{0}}\sum_{s}\frac{\omega_{f}^{2}}{c^{2}}\mathrm{Im}\overleftrightarrow{G}\left(\mathbf{r},\mathbf{r}_{s};\omega_{f}\right)\cdot\mathbf{d}_{s}^{*}\nonumber \\
 & \times\hat{\sigma}_{s}^{12}\left(t\right)\left(\pi\delta\left(\omega_{f}-\omega_{s}\right)+i\mathcal{P}\frac{1}{\omega_{s}-\omega_{f}}\right).
\end{align}
Here, again, we have utilized the relation (\ref{eq:identity}). Using
the above expression and Eq. (\ref{eq:KKrelation}), we obtain the
following expression for the electric field operator 

\begin{equation}
\hat{\mathbf{E}}\left(\mathbf{r},t\right)=\frac{1}{\epsilon_{0}}\sum_{s}\frac{\omega_{s}^{2}}{c^{2}}\overleftrightarrow{G}\left(\mathbf{r},\mathbf{r}_{s};\omega_{s}\right)\cdot\mathbf{d}_{s}^{*}\hat{\sigma}_{s}^{12}\left(t\right).
\end{equation}

Applying the conjugation to the above equation, we can also obtain
the expression for the conjugated field operators $\hat{\mathbf{E}}^{\dagger}\left(\mathbf{r},\tau\right)$.
Inserting these results to Eq. (\ref{eq:spectrum}), we can rewrite
the spectrum as 

\begin{equation}
\frac{dW}{d\Omega}\left(t\right)\approx\sum_{s,s'=1}^{N} K_{ss'} \mathrm{tr}\left\{ \hat{\sigma}_{s'}^{21}\left(t\right)\hat{\sigma}_{s}^{12}\left(t\right)\hat{\rho}\right\} .\label{eq:spectrum-bm}
\end{equation}
with the propagation factors 
\begin{equation}
K_{ss'}=\frac{cr^{2}}{4\pi^{2}\epsilon_{0}}\left[\frac{\omega_{s'}^{2}}{c^{2}}\overleftrightarrow{G}^{*}\left(\mathbf{r},\mathbf{r}_{s'};\omega_{s'}\right)\cdot\mathbf{d}_{s'}\right]\cdot\left[\frac{\omega_{s'}^{2}}{c^{2}}\overleftrightarrow{G}\left(\mathbf{r},\mathbf{r}_{s};\omega_{s}\right)\cdot\mathbf{d}_{s}^{*}\right].
\end{equation}

To compute the spectrum with Eq. (\ref{eq:spectrum-bm}), we need
to evaluate the expectation values $\mathrm{tr}\left\{ \hat{\sigma}_{s'}^{21}\left(t\right)\hat{\sigma}_{s}^{12}\left(t\right)\hat{\rho}\right\} $. To compute these quantities, we consider a pure quantum system. In this case, we can introduce the time-propagation operator $U\left(t\right)$ to reformulate these quantities as
\begin{align}
 & \mathrm{tr}\left\{ \hat{\sigma}_{s'}^{21}\left(t\right)\hat{\sigma}_{s}^{12}\left(t\right)\hat{\rho}\right\} =\mathrm{tr}\left\{U^{\dagger}\left(t\right) \hat{\sigma}_{s'}^{21}U\left(t\right)U^{\dagger}\left(t\right)\hat{\sigma}_{s}^{12}U\left(t\right)\hat{\rho}\right\} \nonumber \\
 & =\mathrm{tr}\left\{\hat{\sigma}_{s'}^{21} \hat{\sigma}_{s}^{12}U\left(t\right)\hat{\rho}U^{\dagger}\left(t\right)\right\} =\mathrm{tr}\left\{\hat{\sigma}_{s'}^{21} \hat{\sigma}_{s}^{12}\hat{\rho}\left(t\right)\right\}.
\end{align}
In essence, we have transformed the expression in the Heisenberg picture to that in the Schr{\"o}dinger picture. To deal with the quantum system in the presence of loss, we should replace $U\left(t\right)...U^{\dagger}\left(t\right)$ as the time-propagation super-operator $\mathcal{U}\left(t\right)$,
which indicates the formal solution of the master equation with loss,
such as master equation (\ref{eq:emq}) or (\ref{eq:master-equation}).
Finally, we can compute the spectrum as 
\begin{equation}
\frac{dW}{d\Omega}\left(t\right)\approx \sum_{s,s'}K_{ss'}\mathrm{tr}\left\{\hat{\sigma}_{s'}^{21}\hat{\sigma}_{s}^{12}\hat{\rho}\left(t\right)\right\}.
\end{equation}

\begin{figure}
\begin{centering}
\includegraphics[scale=0.44]{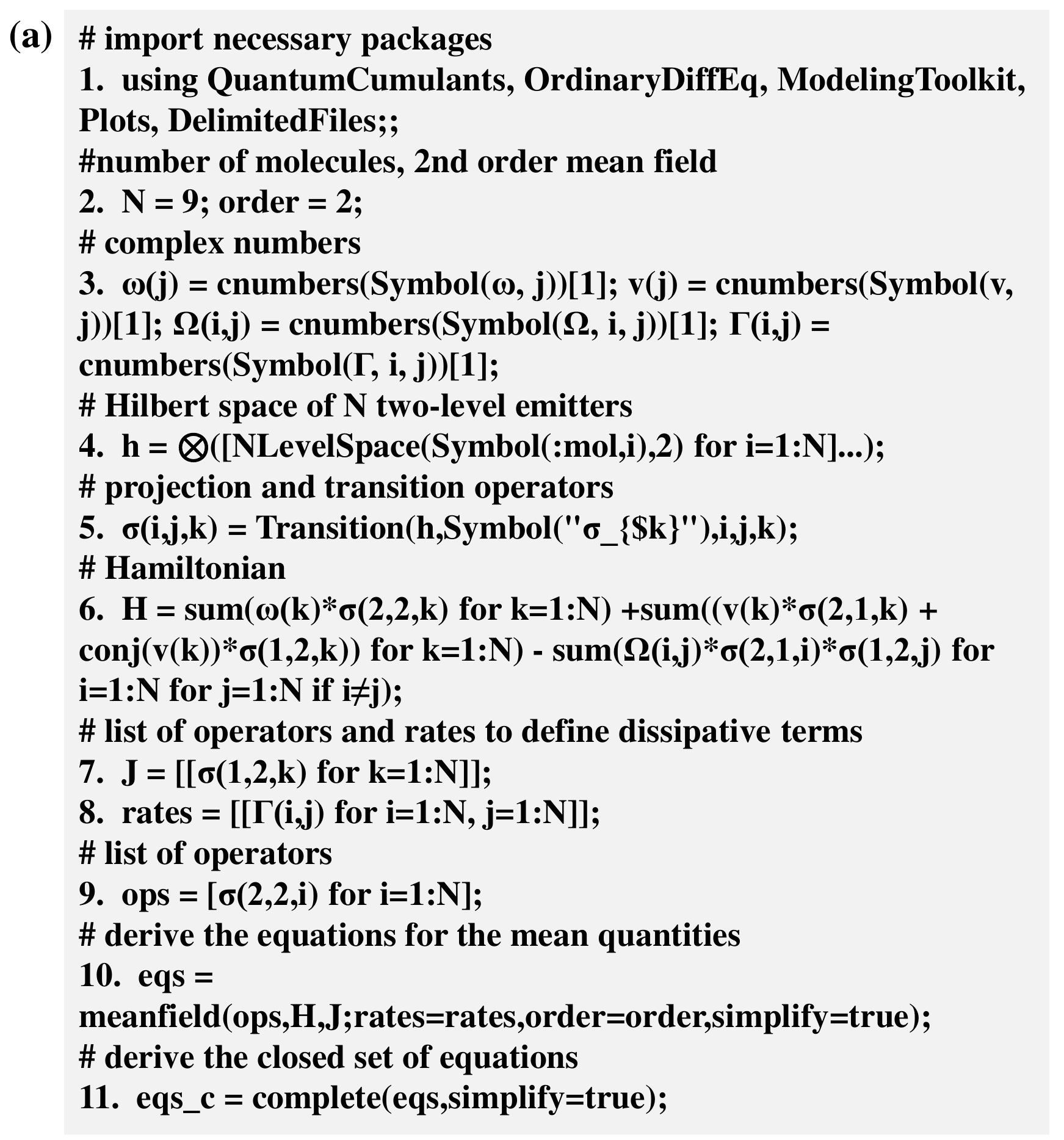}
\includegraphics[scale=0.43]{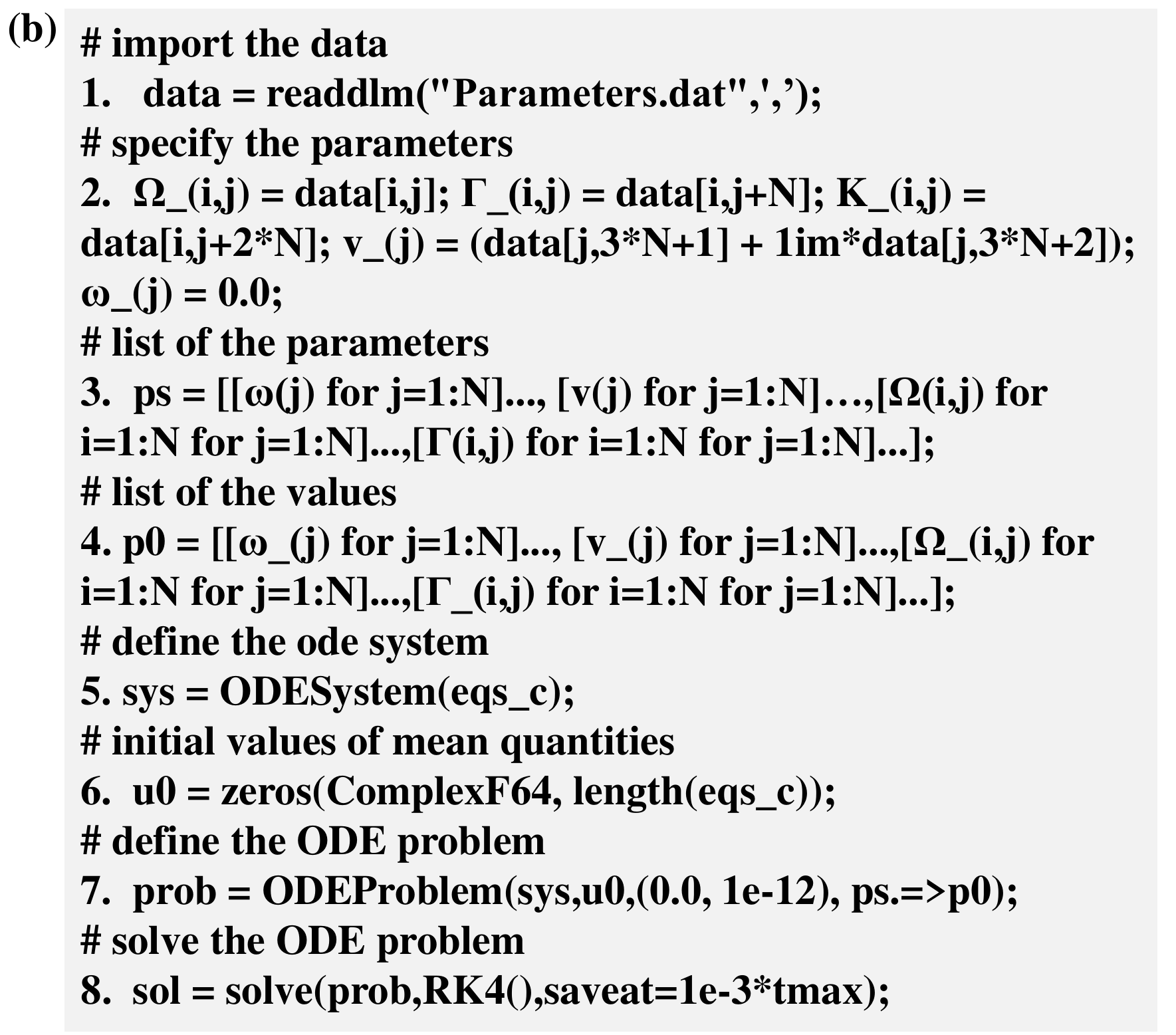}
\includegraphics[scale=0.43]{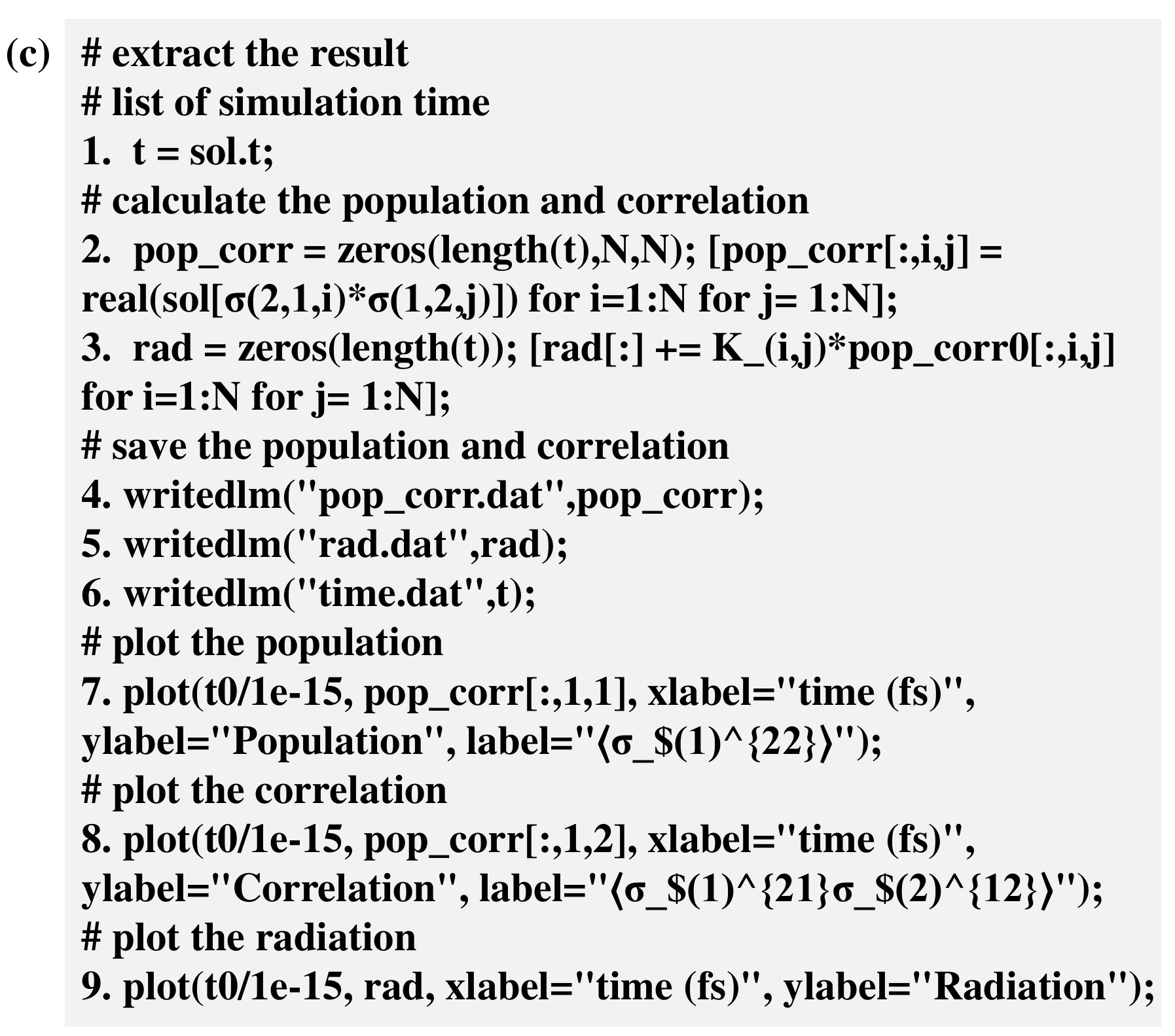}
\par\end{centering}
\caption{\label{fig:code} Panel (a) shows the Julia code to derive the mean-field equations. Panel (b) shows the Julia code to solve the equations. Panel (c) shows the Julia code to extract the numerical results. }
\end{figure}

\section{Julia Codes to Derive and Solve Mean-field Equations\label{sec:codes}}

In this Appendix, we explain the mean-field approach, and the codes to derive and solve the mean-field equations. In this  approach, we derive the equation $\partial_{t}\left\langle \hat{o}\right\rangle =\mathrm{tr}\left\{ \partial_{t}\hat{\rho} \hat{o}\right\} $
for the expectation value $\left\langle \hat{o}\right\rangle =\mathrm{tr}\left\{ \hat{\rho} \hat{o}\right\} $ of
any operator $\hat{o}$, and truncate the resulted hierarchy of  equations with third-order cumulant expansion approximation. The derived equations include those for first-order mean values, e.g. the upper-state populations $\langle\hat{\sigma}_{s}^{22}\rangle$ and the molecular coherences $\langle\hat{\sigma}_{s}^{12}\rangle$, and second-order ones, e.g. the molecular correlations $\langle\hat{\sigma}_{s}^{21}\hat{\sigma}_{s'}^{12}\rangle,\langle\hat{\sigma}_{s}^{22}\hat{\sigma}_{s'}^{22}\rangle,\langle\hat{\sigma}_{s}^{22}\hat{\sigma}_{s'}^{12}\rangle,\langle\hat{\sigma}_{s}^{12}\hat{\sigma}_{s'}^{12}\rangle$.

In the following, we present the Julia code to derive and solve the mean-field equations, see Fig. \ref{fig:code}. First, we explain shortly the code to derive the equations, see Fig. \ref{fig:code} (a). The 1st line imports the necessary packages, and the 2nd line defines the number of molecules and the order of mean-field approach. The 3rd line defines the complex numbers, and the 4th line defines the Hilbert space for the molecules. The 5th line defines the projection and transition operators, and the 6th line defines the system Hamiltonian. The 7th and 8th line define the list of operators and rates to specify the dissipative terms in the master equation. The 9th line defines the list of operators, and the 10th line derives the equations for the expectation values of these operators. The 11th line derives the closed set of the mean-field equations. 

Second, we explain the codes to solve the mean-field equations, see Fig. \ref{fig:code} (a). The 1st line imports the parameters calculated with the MNPBEM, and the 2nd line associates these parameters with those used in the Julia codes. The 3rd and 4th line define the list of the parameters and their values. The 5th line defines  the Ordinary Differential Equations (ODE) system, and the 6th line defines the initial values of the mean-field quantities. The 7th line defines the ODE problem, and the 8th line solves the ODE problem. 

Finally, we discuss the codes to the extract the numerical results, see  Fig. \ref{fig:code} (c). The first line extracts the list of simulation time, and the 2nd line withdraws the population and the correlation. The 3rd line calculates the radiation in the far field, and the 4th to 6th lines save the data. The 7th to 9th lines plot the population, the correlation and the radiation, respectively.

\section{Expectation Values of Collective Spin Operators\label{sec:meanequations}}

In the following, we present the exact expression for the components of the  collective spin vector ${\bf A}=\sum_{i=x,y,z}A_i {\bf e}_{i}$:
\begin{align}
A_x &= \langle\hat{j}_{x}\rangle = (1/2)\sum_{s}\left(\langle\hat{\sigma}_{s}^{12}\rangle+\langle\hat{\sigma}_{s}^{21}\rangle\right), \nonumber \\
A_y &= \langle\hat{j}_{y}\rangle = (i/2)\sum_{s}\left(\langle\hat{\sigma}_{s}^{12}\rangle-\langle\hat{\sigma}_{s}^{21}\rangle\right), \nonumber \\
A_z &= \langle\hat{j}_{z}\rangle = (1/2)\sum_{s}\left(2\langle\hat{\sigma}_{s}^{22}\rangle-1\right), \nonumber 
\end{align}
and the average of the Dicke state quantum numbers $J=\sqrt{\sum_{i}\langle\hat{j}_{i}^2\rangle}$, $M=\langle\hat{j}_{z}\rangle$ with
\begin{align}
\langle\hat{j}_{x}^2\rangle &=\frac{1}{4}\sum_{s,s'}(\langle\hat{\sigma}_{s}^{12}\hat{\sigma}_{s'}^{12}\rangle+\langle\hat{\sigma}_{s}^{12}\hat{\sigma}_{s'}^{21}\rangle+\langle\hat{\sigma}_{s}^{21}\hat{\sigma}_{s'}^{12}\rangle+\langle\hat{\sigma}_{s}^{21}\hat{\sigma}_{s'}^{21}\rangle), \nonumber \\
\langle\hat{j}_{y}^2\rangle &=-\frac{1}{4}\sum_{s,s'}(\langle\hat{\sigma}_{s}^{12}\hat{\sigma}_{s'}^{12}\rangle-\langle\hat{\sigma}_{s}^{12}\hat{\sigma}_{s'}^{21}\rangle-\langle\hat{\sigma}_{s}^{21}\hat{\sigma}_{s'}^{12}\rangle+\langle\hat{\sigma}_{s}^{21}\hat{\sigma}_{s'}^{21}\rangle), \nonumber \\
\langle\hat{j}_{z}^2\rangle &=\frac{1}{4}\sum_{s,s'}(\langle\hat{\sigma}_{s}^{22}\hat{\sigma}_{s'}^{22}\rangle-\langle\hat{\sigma}_{s}^{22}\hat{\sigma}_{s'}^{11}\rangle-\langle\hat{\sigma}_{s}^{11}\hat{\sigma}_{s'}^{22}\rangle+\langle\hat{\sigma}_{s}^{11}\hat{\sigma}_{s'}^{11}\rangle). \nonumber \\
\langle \hat{j}_{z}^2\rangle &=\frac{1}{4}\sum_{s,s'}[4\langle \hat{\sigma}_{s}^{22}\hat{\sigma}_{s'}^{22}\rangle-2(\langle\hat{\sigma}_{s}^{22}\rangle+\langle\hat{\sigma}_{s'}^{22}\rangle)+1]
\end{align}

\section{Supplemental Numerical Results}

In this Appendix, we provide the extra numerical calculations to facilitate the discussions in  the main text.

\begin{figure}[!htp]
\begin{centering}
\includegraphics[scale=0.30]{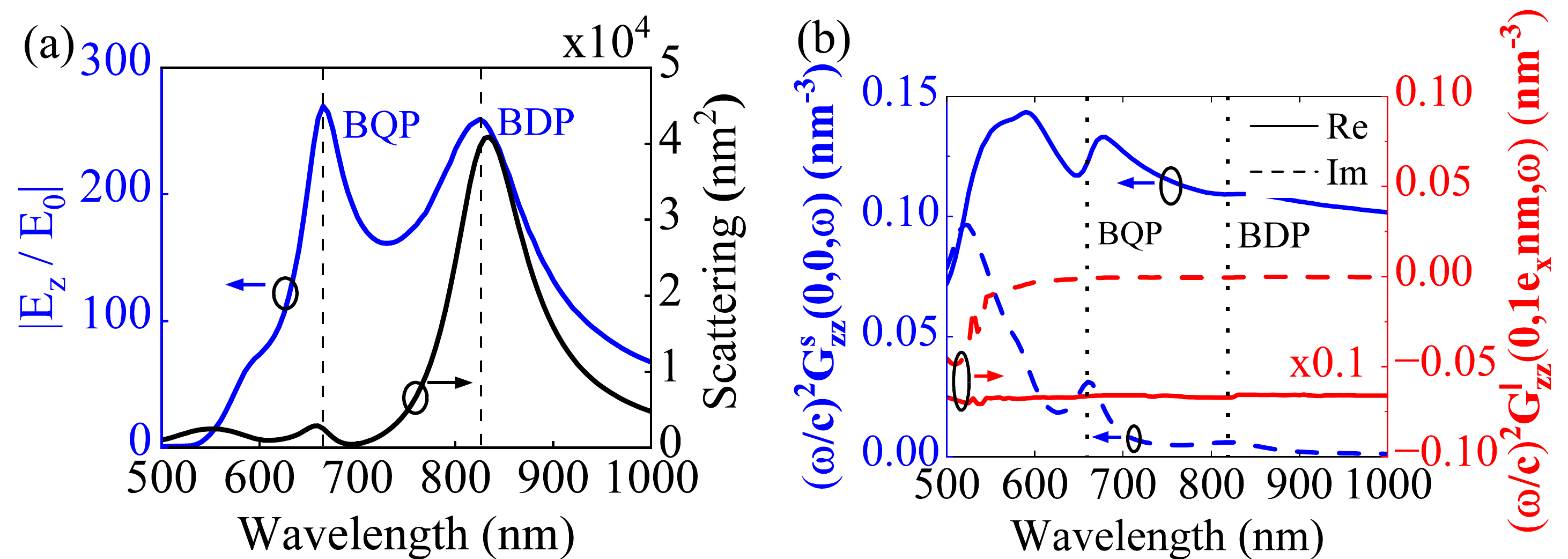}
\par\end{centering}
\caption{ \label{fig:plasmon} Plasmonic response of the metallic nanocavity. Panel (a) shows the scattering cross-section (black solid line, right axis), the enhancement $|E_z(0)/E_0|$ of the z-component electric field (blue solid line, left axis) at the  nanocavity center as function of  light wavelength. The peaks at $660$ nm and $820$ nm are attributed to  the bonding quadruple plasmon (BQP) and the bonding dipolar plasmon (BDP), respectively.  Panel (b) shows the real and imaginary  part  of the scattered Green's tensor zz-component  ${\rm Re}G^s_{zz}(0,0)$, ${\rm Im}G^s_{zz}(0,0)$ (blue solid and dashed line, left axis), and those ${\rm Re}G^l_{zz}(0,{\bf  e}_x)$,${\rm Im}G^l_{zz}(,{\bf  e}_x)$ of the layered ones (red solid and dashed lines), where the red solid line is scaled by $0.1$. For more details, see the text. }
\end{figure}

\subsection{Plasmonic Response of NPoM Nanocavity\label{sec:plasmon}}
We utilize the  boundary element method (BEM)~\citep{FJGDAbajo,FJGDAbajo1}, as implemented in the metal nanoparticle BEM toolkit ~\citep{UHohenester,JWaxenegger}, and together with the  dielectric permittivity of gold as determined in the experiment~\citep{PBJohnson}, and of $2.1$ for the nanogag, to carry out the electromagnetic simulations [Fig. \ref{fig:plasmon}] for the NPoM nanocavity shown in~Fig. \ref{fig:system}. Firstly, we illuminate the nanostructure by a plane-wave with the  polarization and propagation (about $55$ degree to the normal of substrate), and calculate the scattering cross-section, and the enhancement of the electric field along the vertical direction at the middle of nanocavity as function of the light wavelength [Fig.~\ref{fig:plasmon}(a)]. The scattering spectrum shows three peaks at around $820$ nm, $660$ nm, and $520$ nm, which can be attributed to the bonding dipole plasmon (BDP) and the bonding quadruple plasmon (BQP)~\citep{FBenz}, and the transverse plasmon mode. The field enhancement shows two peaks of about $250$ at the BDP and BQP wavelengths, but also one shoulder at around $600$ nm, which can be attributed to the high-order radiative mode~\citep{NKongsuwan}. 

Secondly, we introduce a vertical point dipole in the middle of nanocavity, and then calculate the field scattered from the truncated nanoparticle at the position of the dipole, and the field associated with the corresponding layer structure at a position about $1$ nm away from the dipole. By evaluating the ratio of the fields and the dipole amplitude, we can calculate the scattered and layered dyadic Green's tensor $(\omega/c
)^2G^{s}_{zz}(0,0;\omega)$ and $(\omega/c
)^2G^{l}_{zz}(0,{\bf r} = 1 {\rm nm}{\bf e}_x;\omega)$ [Fig.~\ref{fig:plasmon}(b)]. This particular decomposition was assumed in the MNPBEM~\citep{JWaxenegger} in order to utilize the analytical expressions of the Green's tensor for the layered structures~\citep{MPaulus}. Note that the latter can be further decomposed as the contributions from the propagating field in the free space and the one reflected from the interfaces of the layered structure. Here, we do not consider the layered Green function at the position of the dipole because the free-space field contribution diverges, and consider the $zz$-components because they are relevant for the vertically standing molecules. 

For the former quantity, the imaginary part shows two peaks at around the BDP and BQP wavelengths on the wing of a broad peak  around $520$ nm, which can be attributed to the plasmonic pseudo-mode~\citep{ADelga} formed by the overlapped higher order plasmons. In contrast, the real part shows Fano features around the BDP and BQP wavelengths on a smooth background with positive value. The general feature of the real and imaginary part resembles that of a corresponding metal-insulator-metal structure, and thus can be attributed to the propagating surface plasmon modes~\citep{YZhang2021}. In  contrast, for the latter quantity, the negative real part does not show obvious light-wavelength dependence, which is due to the free-space field, and the negative imaginary part shows feature below $600$ nm with relatively weak value, which is due to the field reflected from the interfaces of layered structure.

\begin{figure}
\begin{centering}
\includegraphics[scale=0.34]{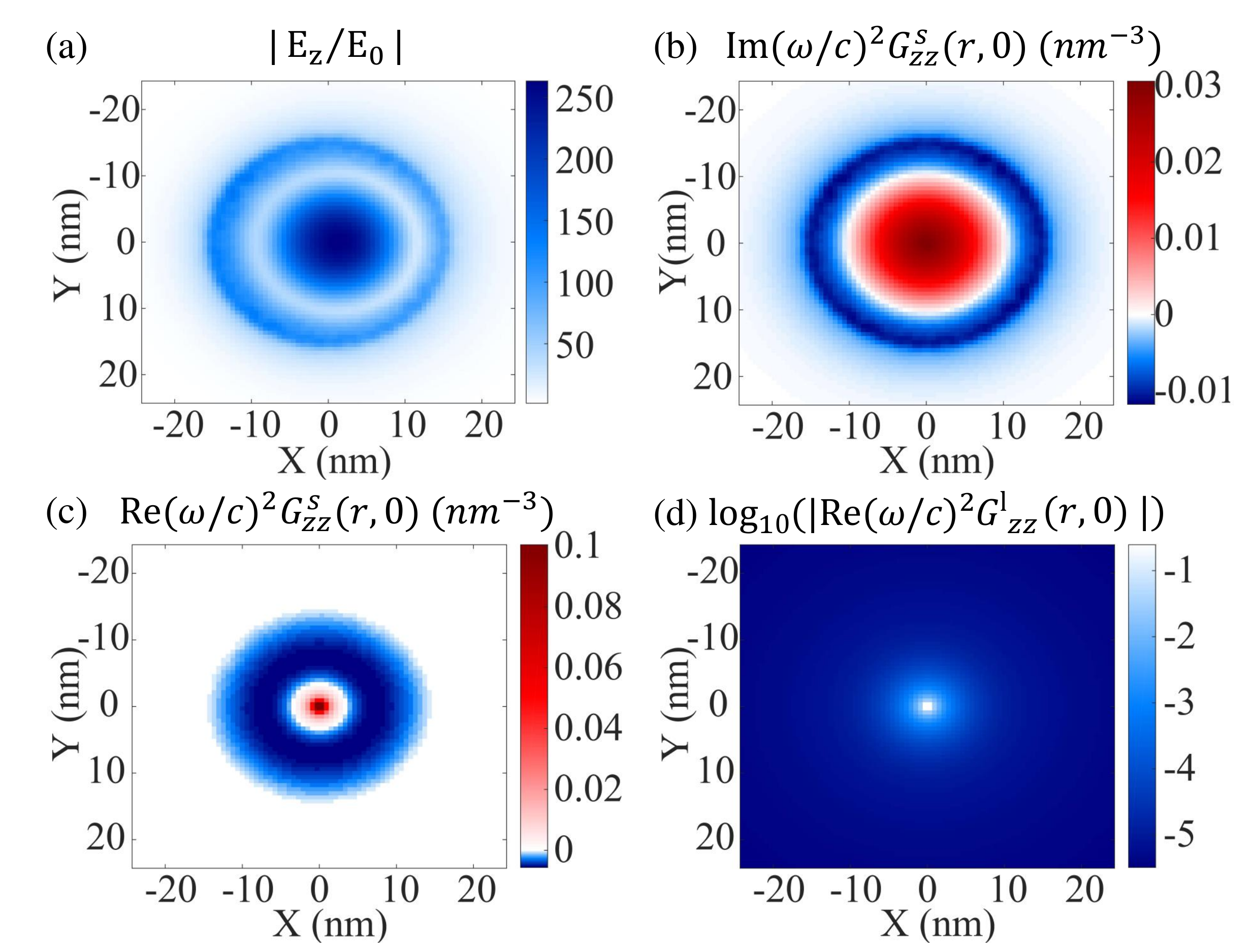}
\par\end{centering}
\caption{\label{fig:spatial-BQP} Spatial dependence of the plasmonic response. Panel (a-d) show the mapping of the electric field enhancement $|E_z({\bf r})/E_0|$ (a), the imaginary part  ${\rm Im}G^s_{zz}(0,{\bf r})$ (b) and real part ${\rm Re}G^s_{zz}(0,{\bf r})$ (c) of the scattered dyadic Green's tensor zz-component, and the real part (d) of the layered dyadic Green's tensor zz-component, at the BQP mode wavelength.}
\end{figure}


In Fig.~\ref{fig:spatial-BQP}, we show the spatial dependence of the plasmonic response at the BQP mode wavelength. We see that the enhancement of the electric field z-component  shows a spot with radius of about $10$ nm, and a ring on the edge of nanocavity [Fig.~\ref{fig:spatial-BQP}(a)], and the imaginary part of the scattered dyadic Green's tensor zz-component shows similar pattern [Fig.~\ref{fig:spatial-BQP}(b)]. In contrast, the real part of the scattered dyadic Green's tensor zz-component shows small spot with radius smaller than $1$ nm over a broad spot with the radius of about $10$ nm, which can be attributed to the short-ranged guiding plasmon mode and the long-ranged gap plasmons. At the same time, the real part of the free-space dyadic Green's tensor zz-component shows only a small spot near the origin.

\begin{figure}
\begin{centering}
\includegraphics[scale=0.34]{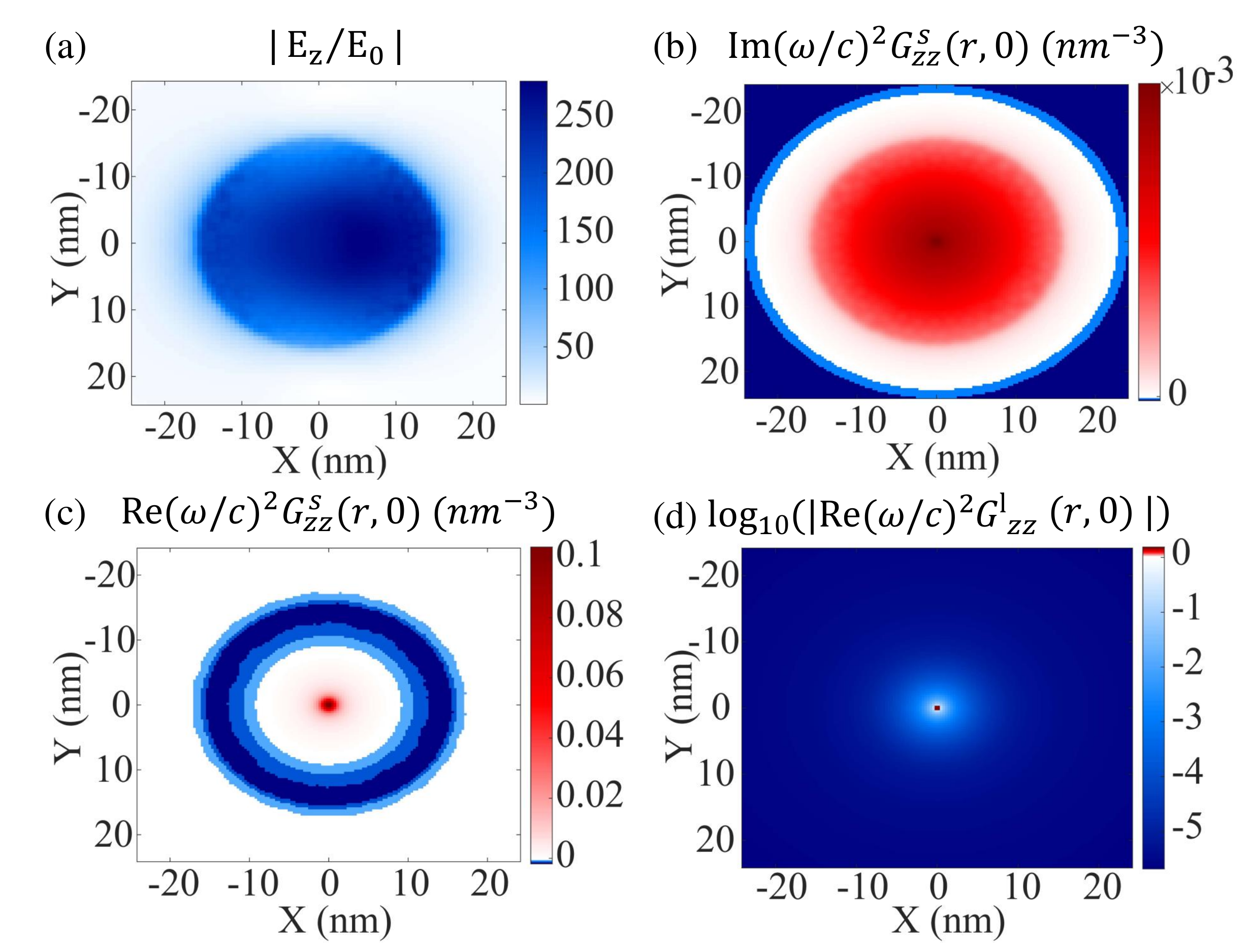}
\par\end{centering}
\caption{\label{fig:spatial-BDP} Similar results as Fig.~\ref{fig:spatial-BQP} but for the BDP mode wavelength.  }
\end{figure}


In Fig.~\ref{fig:spatial-BDP}, we show the spatial dependence of the plasmonic response at the BDP mode wavelength. Fig.~\ref{fig:spatial-BDP} shows similar feature as Fig.~\ref{fig:spatial-BQP} for the BQP mode wavelength except that the center spots with about $10$ nm radius are replaced by a spot through the whole nanocavity. Furthermore, Fig.~\ref{fig:spatial-BDP}(e) and (f) show similar results as Fig.~\ref{fig:plasmon}, which suggests the similar behavior for the molecules resonant to the BQP and BDP mode.

\begin{figure}
\begin{centering}
\includegraphics[scale=0.30]{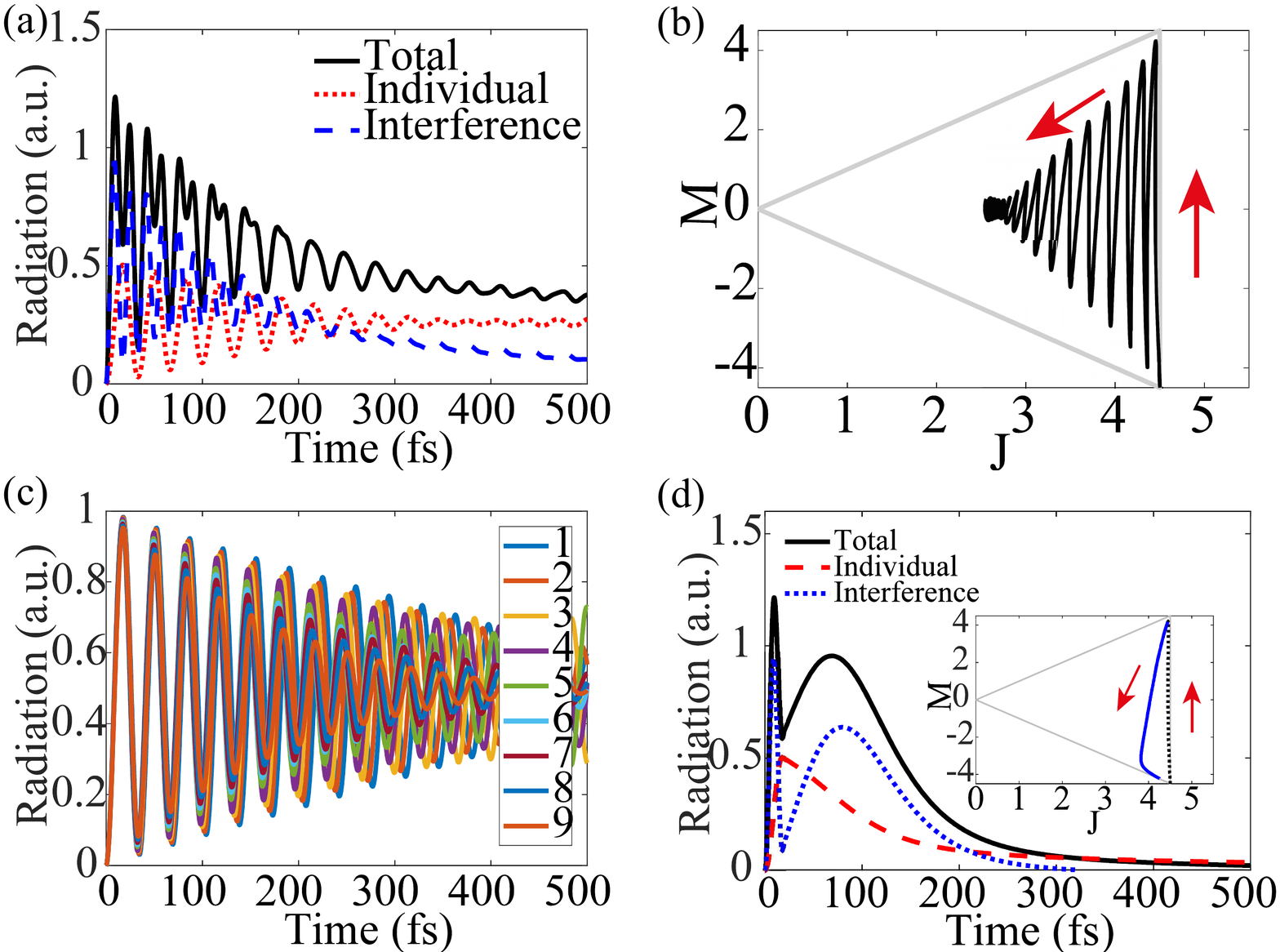}
\par\end{centering}
\caption{\label{fig:sup-linear} Similar results as Fig.~\ref{fig:continuous}(b,c) and Fig.~\ref{fig:pulses} but for a linear array of nine molecules with one nanometer separation inside the nanocavity. Panel (c) shows the population of nine molecules.}
\end{figure}

\subsection{Superradiance for Linear Molecular Array}

In the previous simulations, we have focused on the system with a square array of nine molecules in the middle of the nanocavity center. In this case, the molecules behave similarly due to the similar driving strength, Lamb shift, Purcell-enhanced decay rate, and coherent and dissipative coupling, which is beneficial for the realization of the superradiance. In Fig.~\ref{fig:sup-linear}, we consider a different situation by arranging the nine molecules as a linear array along the radial direction of the nanocavity, so that they experience different parameters according to Fig.~\ref{fig:params}. Fig.~\ref{fig:sup-linear} shows that even in this case the superradiance and the molecular dynamics are similar except that the contribution of the molecular interference to the superradiance is a little reduced, and the resulting superradiant pulse is about three times longer. In addition, in Fig.~\ref{fig:sup-linear}(c), we show that  the excited state populations are initially synchronized, but loss the synchronization at later time, as expected.

\begin{figure}
\begin{centering}
\includegraphics[scale=0.30]{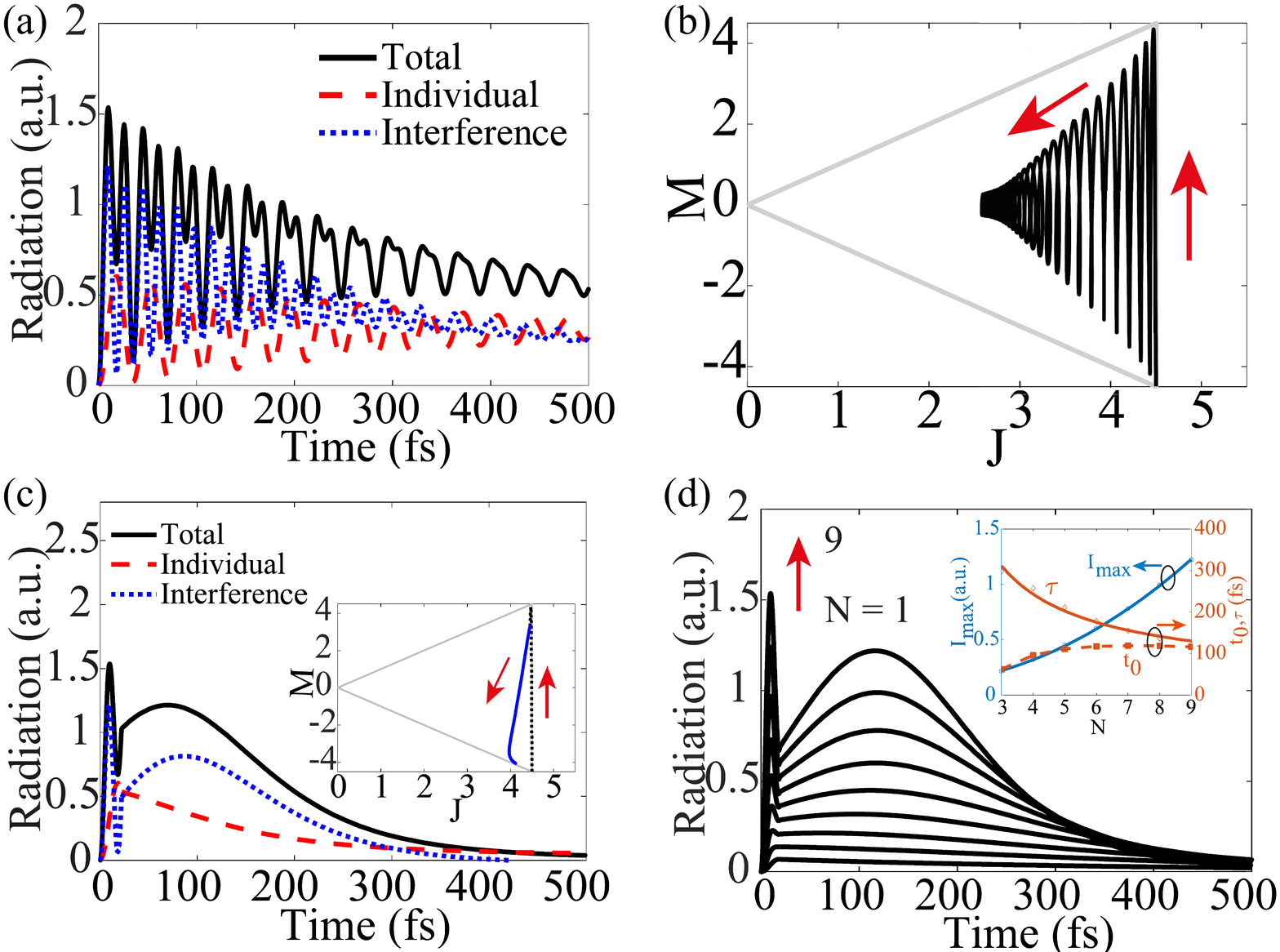}
\par\end{centering}
\caption{\label{fig:sup-BDP}  Similar results as Fig.~\ref{fig:continuous}(b,c) and Fig.~\ref{fig:pulses} for a square array of nine molecules with one nanometer separation. Here, we consider the molecules are resonant to the BDP plasmon with $820$ nm wavelength. In the inset of the panel (d), the pulse maximum and center are fitted as $0.096+0.014N^2$ and $39.73+812.60/N$, respectively.}
\end{figure}

\subsection{Superradiance for Molecules resonant with BDP Plasmon}

In the previous simulations, we focus on a square or linear array of nine molecules in the nanocavity, which are resonant to the BQP plasmon with $660$ nm. In Fig.~\ref{fig:sup-BDP}, we consider a square array of nine molecules, which are however resonant to the BDP plasmon with $820$ nm. In this case, the dissipative coupling between the molecules is much weak, since the imaginary part of the scattered Green's tensor zz-component is much smaller for the BDP plasmon. As a result, we obtain the similar results as shown in Figs.~\ref{fig:continuous} and~\ref{fig:pulses}  except that the continuous superradiance decay much slower, and the superradiant pulses appear much weaker, longer and latter.



\end{document}